\documentclass[onecolumn]{revtex4-1}

\def\be{\begin{equation}}
\def\ee{\end{equation}}
\def\la{\label}
\def\bea{\begin{eqnarray}}
\def\eea{\end{eqnarray}}
\def\non{\nonumber}
\def\fr{\frac}

\def\ci{\cite}
\def\la{\label}
\def\bib{\bibitem}

\def\lm{\lambda}

\def\le{\left}
\def\ri{\right}

\def\p{\phi}
\def\Op{\Omega_{\phi}}
\def\Om{\Omega_{m}}
\def\Opo{\Omega_{\phi o}}
\def\Omo{\Omega_{m o}}
\def\rmm{\rho_m}
\def\rmmo{\rho_{mo}}
\def\rp{\rho_\phi}

\def\G{\Gamma}

\def\Om{\Omega_{m }}
\def\Omo{\Omega_{m o}}

\def\r{\rho}

\usepackage{makeidx}
\usepackage{ctable}
\usepackage{multirow}
\usepackage{graphicx}
\usepackage{grffile}
\usepackage{epstopdf}
\usepackage{subfig}
\usepackage{subfloat}
\usepackage{colortbl}
\usepackage{appendix}
\usepackage{rotating}
\usepackage{array}

\begin{document}

\title {Dark Energy Parametrization motivated by Scalar Field Dynamics}
\author {Axel de la Macorra}
 \email{macorra@fisica.unam.mx}
\affiliation {Instituto de Fisica, Universidad Nacional Autonoma de Mexico,\\  A.P. 20-364, 01000, Mexico D.F., Mexico}

\begin{abstract}

We propose a new Dark Energy parametrization based on the dynamics of a scalar field. We use  an equation of state
$w=(x-1)/(x+1)$, with $x=E_k/V$, the ratio of kinetic energy $E_k=\dot\phi^2/2$ and potential $V$. The eq. of motion gives $x=(L/6)(V/3H^2)$ and has a solution $x=([(1+y)^2+2 L/3]^{1/2}-(1+y))/2$ where $y\equiv \rmm/V$ and  $L\equiv (V'/V)^2 (1+q)^2,\, q\equiv\ddot\p/V'$.
The resulting EoS is $w=\le(6+ L- 6 \sqrt{(1+y)^2+2L/3}\ri)/\le(L+6y\ri)$. Since the universe is accelerating at present time we use the slow roll approximation in which case we have $|q|\ll 1$ and $L\simeq (V'/V)^2$. However, the derivation of  $w$ is exact and has no approximation. By choosing an appropriate ansatz for $L$ we obtain a wide class of behavior for the evolution of Dark Energy without the need to specify the potential $V$. The EoS $w$ can either grow and later decrease, or other way around,  as a function of redshift and it is constraint between $-1\leq w\leq 1$ as for any canonical scalar field with only gravitational interaction.
To determine the dynamics of Dark Energy we calculate the background evolution and its perturbations, since they are important  to discriminate between different DE models. Our parametrization follows closely the dynamics of a scalar field scalar fields and
the function $L$ allow us to connect it  with the potential $V(\p)$ of the scalar field $\p$.

\end{abstract}


\maketitle

\section{Introduction}

In the last years the study of our universe has received a great deal of attention since on the one hand fundamental theoretical questions remain unanswered and on the other hand we have now the opportunity to measure the cosmological parameters with an extraordinary
precision. Existing  observational experiments involve measurement on CMB \ci{planck,wmap9} or large scale structure "LSS" \ci{LSS} or supernovae SNIa \ci{SN}, and new proposals are carried out \ci{new}.

Taking a flat universe dominated at present time by matter  and Dark Energy "DE", and using
a constant equation of state $w$ for DE, one finds   $\Omega_{\rm DE}\simeq 0.714\pm 0.012$,  $\Omega_{\rm m}\simeq 0.286 \pm 0.012 $ with $w=-1.037^{+0.071}_{-0.070}$ from  WMAP9 results \ci{wmap9} and
$\Omega_{\rm DE}\simeq 0.6914^{+0.019}_{-0.021}$,  $\Omega_{\rm m}\simeq 0.208^{+0.019}_{-0.021} $ with $w=-1.13^{+0.13}_{-0.14}$ from  Planck \ci{planck}, SNLS \ci{SN} and BAO \ci{LSS} measurements. The
constraint on curvature is  $-0.0013<\Omega_{k}< 0.0028$ for WMAP9 \ci{wmap9} and $|\Omega_{k}|=0.0005$ for Planck \ci{planck} using a ${\rm \Lambda CDM}$ model, i.e.  $w=-1$ for DE. At present time, the equation of state "EoS"  of DE  depends on the priors, choice of parameters  and on the data used for the analysis as can be seen from the results obtained by either WMAP or Planck collaboration groups  \ci{planck,wmap9}. A more precise determination of the EoS of DE will be carried out in  \ci{new} which together with precise measurements of CMB such as  \ci{planck,wmap9} will yield
a better understanding of the dynamics of Dark Energy. With better data we should be able to study in more detail the nature of Dark Energy, a topic of major interest in the field \ci{DE.rev}. Since the properties of Dark Energy are still under investigation, different DE parametrization have been proposed to help discern on the dynamics of DE \ci{DEparam}-\ci{quint.ax}. Some of these  DE parametrization have the advantage of having a reduced number of parameters, but they may lack a physical motivation and may also be too restrictive.  The evolution of DE background  may not be enough to distinguish between different DE models and therefore the perturbations of DE may be fundamental to differentiate between them.

Perhaps the best physically motivated candidates for Dark Energy are scalar fields, which can interact only via gravity  \ci{SF.Peebles,tracker, quint.ax} or interact weakly  with other fluids, e.g. Interacting Dark Energy "IDE" models\ci{IDE,IDE.ax}. In this work we will concentrate on canonically normalized scalar fields minimally coupled to gravity. Scalar fields have been widely studied in the literature \ci{SF.Peebles,tracker, quint.ax} and special interest was devoted to tracker fields \ci{tracker}, since in this case the behavior of the scalar field $\p$ is weakly dependent on the initial conditions set at an early epoch, well before matter-radiation equality. In this class of models a fundamental question is why DE is relevant now, also called the coincidence problem, and this can be understand by the insensitivity of the late time dynamics on the initial conditions of $\p$. However,  tracker fields may not give the correct phenomenology since the have a large value of $w$ at present time.  We are more interested at this stage  to work from present day redshift  $z=0$ to larger values of $z$, in the region where DE and its perturbations are most relevant. Interesting models for DE and DM have been proposed using gauge groups, similar to QCD in particle physics, and have been studied to understand the nature of Dark Energy \ci{GDE.ax} and also Dark Matter \ci{GDM.ax}.

Here we propose a new DE parametrization based on scalar fields dynamics, but the parametrization of $w$ can be
used without the connection to scalar fields. This parametrization has a rich structure that allows $w$ to have different evolutions, it may grow and later decrease or other way around. We  also determine  the perturbations of DE which together with the evolution of the homogenous part  can single out the nature of DE. With the underlying  connection between the evolution of $w$ and the dynamics of scalar field we could determine the potential $V(\p)$.
The same motivation of parameterizing  the evolution of scalar field was presented  in an interesting paper \ci{huang}. We share the same motivation but we follow a different path. We have the same number of parameters but  a richer structure and it  is easier to obtain information of the scalar potential $V(\p)$ .

We organize the work as follows:  in Sec.\ref{ow} we give a brief overview of our DE parametrization. In  Sec.\ref{sfd} we present the scalar field and we define the variables used in this work. In Sec.\ref{sec.dyn} we present the dynamics of a scalar field  and the set up for our DE parametrization  given in Sec.\ref{sfp}. We calculate the DE perturbations in Sec.\ref{sec.pert} and finally we conclude in Sec.\ref{con}.

\subsection{Overview}\la{ow}

We present here an overview of our  $w$ parametrization. The EoS is
\be\la{wo}
w=\fr{p}{\r}=\fr{x-1}{x+1}.
\ee
with $x\equiv E_k/V$ the ratio of kinetic energy $E_k=\dot\p^2/2$ and potential $V(\p)$. This corresponds to
a canonically normalized scalar field. The
equation of motion of the scalar field gives  (c.f. eq.(\ref{x3})),
\be\la{xo}
x=\fr{\le(\sqrt{1+\fr{2 L}{3(1+y)^2}}-1\ri)\le(1+y\ri)}{2},
 \ee
where $y\equiv \rmm/V$  the ratio of matter density and $V$,  $ L\equiv (V'/V)^2 A$ with $A\equiv (1+q)^2, q\equiv\ddot\p/V'$.
Eq.(\ref{xo}) is an exact equation and is valid for any fluid evolution and/or for an arbitrary potential $V(\p)$.
In terms of $L$ and $y$ the EoS is
\be
w=\fr{6+ L- 6 \sqrt{(1+y)^2+2L/3}}{L+6y}=-1+ \fr{6(1+y)- 6 \sqrt{(1+y)^2+2L/3}}{L+6y}
\ee
which we consider our master equation and it is valid for any value of $L$ and $y$ and not only in the slow roll regime.

The aim of our proposed parametrization for $L,y$  is to cover a wide range of DE behavior. Of course other
interesting parameterizations are possible. The dynamics
of scalar fields  with canonical kinetic terms has an EoS constrained between $-1<w<1$
and  gives  an accelerating universe only if $\lm=-V'/V \rightarrow 0$
or to a constant $|\lm|<1$ with $w=-1+\lm^2/3$ \ci{quint.ax} (for an EoS  $w<-0.8$ one needs $\lm<0.8$).
From the dynamics of scalar fields we know that the evolution of
$w$ close to present time is model dependent. For example, in the case of $V=V_o\p^{-2/3}$, used as a
model  of DE derived from gauge theory \ci{GDE.ax}, the shape of  $w(z)$ close to present time depends on the initial
conditions and it may grow or decrease as a function of redshift $z$.  We also know that tracker fields
are attractor solutions  but in most cases they do not give a negative enough $w_o$ \ci{tracker}. In this class of models the EoS, regardless
of its initial value, goes to a period of kinetic domination where $w\simeq 1$ and later has a steep transition
to $w\simeq -1$, which may be close to present time, and finally it grows to $w_o$ in a very model and initial condition  dependent way. Furthermore,
if  instead of having a single potential term we have two competing terms close to present time, the evolution
of $w(z)$ would even be more complicated. Therefore, instead of deriving the potential $V$ from theoretical
models as in \ci{GDE.ax} we propose to use an ansatz for the functions $L$ and $y$ which on one hand should cover
a wide range of DE behavior with the least number of parameters but without sacrificing generality, and on the other hand we would like
to have this ansatz as close as possible to the known scalar field dynamics. We believe  that using our model will greatly simplify the extraction of DE dynamics from the future observational data. We propose therefore the ansatz (c.f. eqs.(\ref{ll}) and (\ref{f2}) )
 \bea
L=&&L_o+L_1 y^\xi f(z)\\
f(z)=&&\fr{(z/zt)^k}{1+(z/z_t)^k}
\eea
where  $y= y_o a^{3w_o}$ and $L_o,L_1$ are free parameters giving $w_o$ at present time (a subscript $o$ represents present time quantities)
and $w_i=w(z \gg z_t)$  at early times, $f(z)$ is a function that goes from $f(z=0)=0$  at $z=0$  to $f(z\gg 1) =1$ at large $z$. The parameter $z_t $ sets
the transition redshift between $w_o$ and $w_i$  while $k$ sets the steepness of the transition and  $\xi$ takes only two values $\xi=1$ or $\xi=0$ (see sect.(\ref{sfp})).

Since the universe is accelerating at present time we may take the slow roll approximation where $|q|\ll 1, A\simeq 1$ and $L\simeq (V'/V)^2$.
In this case one has $w\simeq -1 + L/3(1+y)$. However, the derivation of $L$ in eq.(\ref{xo}) is exact  and has no approximation. We will show in section \ref{sfp} that $w$ can have a  wide range of behavior and in particular $w$ can decrease and later increase  as a function of redshift
and vice versa, therefore  the shape and steepness are not predetermined by the choice of our parametrization. Of
course we could use other parametrizations for $L$ since the evolution of $x$ and $w$ in eqs.(\ref{wo}) and (\ref{xo})
are  fully valid. There is also no need to have any  reference  to the underlying scalar field dynamics,
i.e. our parametrization is not constrained to scalar field dynamics.
However, it is when we interpret $x,L$ and $y$ as: $x\equiv \dot\p^2/2V$,  $ L= (V'/V)^2 A$ and  $y= \rmm/V$, that we
connect the evolution of $w$ to the scalar potential $V(\p)$.

Finally DE perturbations are important in distinguishing between different DE models \ci{hu}-\ci{DE.p} and we will
show that a steep  transition of $w$ has a bump in the adiabatic sound speed $c_a^2$ which could be
detected in large scale structure \ci{DE.p.sf,DE.p}. At an epoch where the universe is dominated by DM and DE
the total perturbation $\delta_T = \delta\r_T/\r_T= \Omega_{\rm DM} \delta\r_{DM} + \Omega_{\rm DE}  \delta\r_{DE}$ and if  $\Omega_{\rm DE}$ is not much smaller than $\Omega_{\rm DM}$, i.e. for low redshift $z$,  then the evolution of $\delta\r_{DE}$  may have an important contribution to $\delta_T $, as discussed in Sec.\ref{sec.pert}.

\section{Scalar Field Dynamics}\la{sfd}

We are interested in obtaining a new DE parametrization  inferred from scalar fields \ci{SF.Peebles}-\ci{IDE.ax}.
Since it is derived from the dynamics of a scalar field $\p$ we can also determine its
perturbations which are relevant in large scale structure formation.
We start with a FRW metric with a line element
\be
ds^2=dt^2-a(t)^2\le(\fr{dr^2}{1-kr^2}+r^2d\theta^2+r^2\sin(\theta)d\varphi^2 \ri)
\ee
and a canonically normalized  scalar field  $\p(t,x)$ with a potential $V(\p)$ and minimally coupled to gravity.
The homogenous part of $\p$  has an equation of motion
 \be\la{eqm}
 \ddot \p + 3H\dot\p+V'(\phi)=0
 \ee
where  $V'\equiv dV/d\p$, $H=\dot a/a$ is the Hubble constant, $a(t)$ is the scale factor
and a dot represents derivative with respect to time $t$. Since we are interested in the
epoch for small redshift z, with $a_o/a=1+z$, we only need to consider matter and DE and we have
 \be\la{hh}
3H^2=\rmm+\rp,
\ee
in natural units $8\pi G=1, c=\hbar=1$.  The energy density $\r$  and pressure $p$
for the scalar field are
\be\la{rp1}
\rp=\fr{1}{2}\dot\p^2+V ,\;\;\; p_\p=\fr{1}{2}\dot\p^2-V.
\ee
We define  the ratio of kinetic energy and potential energy as
\be\la{x}
x\equiv \fr{\dot\p^2}{2V}
\ee
and the equation of state parameter "EoS" becomes
\be\la{w}
w\equiv \fr{p_\p}{\rp}=\fr{\dot\p^2/2-V}{\dot\p^2/2+V}=\fr{x-1}{x+1}
\ee
or
\be\la{wdw}
w=-1+\delta w, \;\; \;\;\;\; \delta w\equiv \fr{2x}{1+x}.
\ee
The value of $x$ determines $w$ or inverting eq.(\ref{w}) we have
$ x=(1+w)/(1-w)$. Since  $x \geq 0  $  the Dark Energy  EoS $w$ is in the range $-1\leq  w\leq 1$.
For growing $x$ the EoS $w$ becomes larger and at $x\gg 1$ one has $w\simeq 1$ while a decreasing $x$ has
$w \rightarrow -1$ for $x \rightarrow0$. In terms of $x$ we have
 \be\la{rpx}
 \rp=V(x+1),\;\; p_\p=V(x-1),\;\;  \rmm=V y
 \ee
where we defined the ratio of $\rm$ and scalar potential $V$ as
\be\la{yd}
y\equiv \fr{\rmm}{V}.
\ee
Finally we can express $\Op,\Om$ and $H$ in terms of $x,y$ and $V$ as
 \bea\la{opm}
\Op&=&\fr{1+x}{1+x+y},\;\; \;\;\Om=\fr{y}{1+x+y}, \\
3H^2&=&\rmm +\rp=V(1+x+y).
\la{h} \eea
The equation of motion for $\p$ is
\be\la{pq}
\dot\p=-\fr{V'+\ddot\p}{3H}=-\fr{V'(1+q)}{3H}
\ee
and we can write
\be\la{x1}
x\equiv \fr{\dot\p^2}{2V}=  \fr{V}{3H^2}\fr{V'^2}{6V^2}\le(1+q\ri)^2=\fr{V}{3H^2}\fr{L }{6}
\ee
with
\bea\la{lq}
L\equiv &&\le(\fr{V'}{V}\ri)^2 (1+q)^2=\lm^2 A,\\
 \lm\equiv &&-\fr{V'}{V}, \;\;\;\; \;\; q\equiv \fr{\ddot\p}{V'},\;\;\;\;\; \;\; A\equiv\le(1+q\ri)^2.  \nonumber
\eea
Since the r.h.s. of eq.(\ref{x1})  still depends on $x$ through $H$ we use
eq.(\ref{h}) and eq.(\ref{x1})  becomes then
 \be\la{xy}
 x=\fr{L }{6(1+x+y)}
 \ee
which has a simple solution
 \be\la{x3}
 x=\fr{\le(\sqrt{1+\fr{2 L }{3(1+y)^2}}-1\ri)\le(1+y\ri)}{2}=\fr{\sqrt{(1+y)^2+2L/3}-(1+y)} {2}.
 \ee
Substituting eq.(\ref{x3}) into eq.(\ref{w}) we obtain our DE parametrization as a function of $L$ and $y$
\be\la{wde}
w =\fr{\sqrt{(1+y)^2+2 L/3}-(3+y)}{\sqrt{(1+y)^2+2L/3}+(1+y)}.
\ee
If we multiply in eq.(\ref{wde}) the numerator and denominator  by  $\sqrt{(1+y)^2+2L/3}-(1+y)$
we obtain an alternative and useful  expression for the EoS
\be\la{www}
w=\fr{6+ L- 6 \sqrt{(1+y)^2+2L/3}}{L+6y}=-1+ \fr{6(1+y)- 6 \sqrt{(1+y)^2+2L/3}}{L+6y}
\ee
and
\be
\delta w=w+1=\fr{6(1+y)- 6 \sqrt{(1+y)^2+2L/3}}{L+6y}
\ee
which we consider our master expression for $w$. It is a relatively simple expression but more important is that
eqs.(\ref{wde}) and (\ref{www}) are exact and therefore valid for any values of $L$ and $y$ and not only at a slow roll regime.
We will see in the next section some physically motivated  limits of eq.(\ref{www}).
Finally, inverting the expression of $w$ we can obtain $L$ as a function of $w$ and $y$
\be\la{lw}
L=\fr{12(1 + w) + 6 y(1 - w^2 )}{(1 - w)^2}
\ee
and $y$  as a function of $w$ and $L$
\be\la{yw}
y=\fr{ L(1 - w)^2-12(1+w)}{6 ( 1- w^2)}.
\ee

\subsubsection{Dynamics and Limits for $L,x,w$}\la{lim}

Using eq.(\ref{xy}) we can express
\be\la{wdw2}
w=-1+\delta w =-1 + \fr{2L}{L+6(1+x+y)},
\ee
and it is an exact equation, however $x$ depends on $y,L$ through eq.(\ref{x3}). If  we
take  $x< (1+y)$, valid in the slow roll approximation,  one has for arbitrary values of  $y$ and $L$
\be\la{lm1}
x=\fr{L}{6 (1 + y)},\;\;\;  w=-1+\fr{L}{3(1+y)}
\ee
and as $L\rightarrow 0$ we have $x\rightarrow 0$ and $w\rightarrow -1$. Notice that
\be\la{yy1}
y=\fr{\rmm}{V}=\fr{\rmm}{\rp}\fr{2}{(1-w)}=\fr{\Om}{\Op}\fr{2}{(1-w)}
\ee
and we then expect that $y$ increases as a function of redshift  z  when
matter dominates over Dark Energy and with $y_o=y(a_o)\simeq 0.25$ at present time.
We expect then that $x< (1+y)$ will be satisfied beyond the slow roll approximation and eq.(\ref{lm1}) is  valid
for a wide range of the parameter's  values.

If we take $L\gg 1$ with $y$ constant in eq.(\ref{www}) we get the limit
\be\la{lm2c}
x= \sqrt{\fr{L}{6}}, \;\;\; w= 1 -2\sqrt{\fr{6}{L}}
\ee
and for  $L/y\rightarrow L_1$ constant with $y\gg 1, L\gg 1$
we have   $x$ and  $w$  also constants given by
\be\la{lm3}
x=\fr{L_1}{6}, \;\;\;  w=\fr{L_1-6}{L_1+6}=-1+\fr{2L_1}{6+L_1}.
\ee
Clearly depending on the value of $L$ we can
have a decreasing or increasing $x$ and $w$ as a function of redshift.
For example for $w=0$ one has $L_1=6$ while for $w=1/3$ one requires $L_1=12$ at large $y$.

Since $w$ is only a  function of $x$ we have $dw=w_x dx$  or as a function of redshift
\be\la{wz}
w_z= w_x x_z
\ee
with
\be
w_x=\fr{2}{(1+x)^2} \geq 0.
\ee
The sign of $w_z$ depends then only on the sign of $x_z$ given by
\be\la{xz}
x_z=x_y y_z + x_L L_z
\ee
with
\be\la{xl}
x_y = - \fr{x}{1+2x+y},\;\;x_L =\fr{1}{6(1+2x+y)}.
\ee
Clearly $x_L$ is positive definite while $x_y\leq 0$.  In general we can assume that DE redshifts slower than matter, at least for small  z ,  since DE has $w<0$ and matter $w_m=0$, so $y=\rmm/V$ is a growing function of z, i.e. $y_z>0$ and $x_y y_z\leq 0$.  Therefore if $L_z$
is negative we have   $x_L L_z <0$ and a decreasing $x$ and $w$ as a function of redshift. However,  if $L_z$ is positive  then the sign of
$x_L L_z$ is positive and $w$ may grow or decrease depending on the magnitude of  $|x_y y_z|$ compared  to $|x_L L_z|$.

\subsubsection{Slow Roll Approximation}

The evolution of $x$  as a function of $L$ and $y$ is given in  eq.(\ref{x3}) and
here we will show some phenomenological interesting limits.
Since the field $\p$ should be  responsible for accelerating the universe we know
that $w$ must be close to $-1$ at present time, and the field $\p$ must satisfy the slow roll approximation.
In the slow roll approximation one has $|\ddot\p|\ll |3H\dot\p| \simeq |-V'|$ and we  then  have
\be
|q\equiv \fr{\ddot\p}{V'}|=|\fr{3H\dot\p}{V'}-1| \ll 1
\ee
with  $A\simeq 1 $ and the function $L$ becomes
\be
L\simeq\le(\fr{V'}{V}\ri)^2 =\lm^2.
\ee
We name  a full slow roll approximation when $\ddot\p=0$  (i.e. $3H\dot\p=-V'$ and  $q=0$). This full slow roll approximation is more suitable in inflation where
one has a long  period of inflation, however Dark Energy does dominate only recently and so we do not expect
this approximation to hold for a long period of time and should not be taken as a working hypothesis for DE parametrization.

\subsubsection{Late time attractor Solution}

The evolution of scalar fields has been studied in \ci{quint.ax} and a late time attractor for
an accelerating universe  requires
$w < -1/3, \lm^2 < 2 $ with  $\Op \rightarrow 1$. In the limit $\lm^2 < 3$ with $\rmm\ll \rp$ one has
\be\la{at1}
\fr{\dot\phi^2}{6H^2}=\fr{\lm^2}{6}, \;\;\;\;\;\; \fr{V}{3H^2}=1-\fr{\lm^2}{6}
\ee
giving
\be\la{at2}
x\equiv \fr{\dot\p^2}{2V} = \fr{\lm^2}{6-\lm^2}, \;\;\;\;\;\; w= -1+\fr{\lm^2}{3}.
\ee
Notice that  eq.(\ref{lm1}) reduces to  eq.(\ref{at2}) in the limit $y\ll 1$  with $L\simeq \lm^2$, and
therefore  eq.(\ref{lm1}) generalizes  eq.(\ref{at2}). For large redhsift z we expect $y$ to increase and eq.(\ref{at2})
would not longer be valid and we should take instead eq.(\ref{lm1}).

\section{Dynamical Evolution  }\la{sec.dyn}

Differentiating $x$ and $y$   w.r.t. time or
equivalently  as a function $N\equiv \rm ln[a]$, where $a$ is the scale factor,
we get the evolution of $x$ and $y$. Using  the definition of $x$ in eq.(\ref{x1}),
$\dot\rmm=-3H\rmm$, $\ddot\p=-3H\dot\p-V'$  and   $f_N\equiv df/dN= \dot f/H $ for any
function $f[N(t)]$, we get
\be
\dot x = \fr{\dot\p}{V}\le(\ddot\p-x V'\ri)=\fr{\dot\p V'}{V}\le(q-x\ri) =6Hx\le(\fr{x-q}{1+q}\ri)
\ee
\be\la{dxy}
 x_N=   \fr{\dot x}{H}=6x\le(\fr{x-q}{1+q}\ri)=6x\le(\fr{1+x}{1+q}-1\ri)
\la{dxx}\ee
and $\dot y = H y_N$ with
\be\la{dy}
y_N= \fr{\dot y}{H}=  -3y - \fr{yV'\dot\p}{H V}=  -3y\le(1-\fr{2x}{1+q}\ri).
\ee
We see that eqs.(\ref{dxy}) and (\ref{dy}) are uniquely determined
by a single function $q\equiv\ddot\p/V'$.
The critical points for $\dot x=0$, i.e. $\dot\p=0$, have  $x=0$  or $x=q$
while $\dot y=0$ is satisfied for $y=0$ and $2x=1+q$.
The case, $\dot\p=x=0$ has $ w=-1$  and
$\dot y= -3Hy$ which gives a solution $y=y_i(a/a_i)^{-3}$,  a constant $V(\phi)$  and the limits
$\Op \rightarrow 1, y \rightarrow 0$  and $\Om \rightarrow 0$.
In the case $q=x$ the EoS  becomes $w=(q-1)/(q+1)$ and it will take different
constant values. Setting $q=x$ constant in eq.(\ref{dy}) we get $\dot y=3Hyw$  with $w$ constant
one has a solution
\be\la{yss}
y=y_o\le(\fr{a}{a_o}\ri)^{3w}.
\ee
For $q=x<1$ we have  $w<0$, $y \rightarrow 0$ and $ \Op \rightarrow 1$.
Finally, we can
satisfy $\dot x=\dot y=0$  for $x=y=0$ or $x=q=1$ giving $w=0$ and  $y$ constant (c.f. eq.(\ref{yss})).

Therefore, having an increasing or decreasing $x$ depends on the sign of
$\dot\p(\ddot\p-x V')=\dot V(q-x)$ and it can vary as a function on time depending on the values of $x$ and $q$, i.e.
on the choice of the potential $V(\p)$.  If we take what we call a full slow roll defined by  $\ddot \p=q=0$ and $3H\dot\p=-V'$ then eq.(\ref{dxy}) becomes
\be\la{dx2}
\dot x= 6Hx^2
\ee
which is positive definite,  i.e. $\dot x \geq 0$. Using $H= \dot a/a$
we can express eq.(\ref{dx2})  giving a  solution
\be\la{sx}
x=\fr{x_o}{1+6 x_o Ln(a_o/a)}=\fr{x_o}{1+6 x_o Ln(1+z)}.
\ee
Therefore if the condition $q=0$ or $|q=\ddot \p/V' |\ll x$ is satisfied, then from eq.(\ref{sx}) we have
a decreasing function for $x$ as a function of redshift z and therefore $w(z)$ also decreases. However, we
do not expect the universe to be in a full slow roll regime and
when $x$ is small, e.g. $w<-0.9$ one has $x<0.05$, the slow roll condition
$|\ddot \p| <  |V'|$ does not imply that   $q=\ddot \p/V' \ll x $. Therefore, the sign of $\dot x$ can be positive or negative
depending on the sign and size of $q=\ddot \p /V'$ compared to $x$ and $x$ can either  grow or decrease.
In the region where  $q=\ddot \p/V'< x $ we have  $\dot x>0$ while for $q=\ddot\p/V' > x $ we have $\dot x<0$.
The value of $q$ parameterizes then the amount of slow roll of the potential and a full slow roll has
$q=0$ but we expect to be only in an approximate  slow roll regime  with $|q|\ll 1$ and $L\simeq \lm^2$.
We discuss the dynamics  of $q$ in section \ref{sec.pert}.
In the present work we do not want to study the critical points of the dynamical equations but the evolution of $x$ close
to present time when the universe
is accelerating with $x$ close to zero ($w$ close to -1) but not exactly zero  with $\dot\p\neq 0$ and $\ddot\phi\neq xV'$

\subsection{Dynamical evolution  of $x$ and $y$}

The dynamical eqs.(\ref{dxx}) and (\ref{dy}) can also be written in terms of $V_N/V, \,V_N\equiv dV/dN$ as
\bea\la{dxyn}
x_N &\equiv & \fr{dx}{dN} = - \le(6x+ (1+x)\fr{V_N}{V}\ri)\nonumber\\
y_N &\equiv & \fr{dy}{dN} =  -y\le(3+\fr{V_N}{V}\ri).
\eea
From eqs.(\ref{dxyn}) we clearly see that the behavior of $x$ and $y$ is completely determined by a single function $V_N/V$
and we define
\be\la{zz}
\zeta\equiv - \fr{V_N}{V}=-\fr{V'\p_N}{V}=\lm(\p)\;\p_N
\ee
with $\lm=-V'/V$. It is easy to see from eqs.(\ref{dxy}), (\ref{dy}) and (\ref{dxyn}) that the function $\zeta$
is given in terms of $x$ and $q$ as
\be\la{zq}
\zeta= - \fr{V_N}{V}=\fr{6x}{1+q}.
\ee
In the case where the scalar field rolls down its minimum we will have $\dot V=V'\p_N$ negative giving a $\zeta > 0$.
In the simple case of having  $\zeta$ constant the solution to eqs.(\ref{dxyn}) are simply
\be\la{xs1}
x=x_i e^{-(6-\zeta)(N-N_i)}+ \le(\fr{\zeta}{6-\zeta}\ri)\le(1-e^{-(6-\zeta)(N-N_i)}\ri)
\ee
and
\be\la{ys1}
y=y_i e^{-(3-\zeta)(N-N_i)}
\ee
with initial conditions $x=x_i, y=y_i$ at $N=N_i$. The asymptotic values of $x$ and $y$ depend
on the value of $\zeta$.  The critical points of the system  in eqs.(\ref{dxyn})  give for
$x_N=0$ a solution
\be\la{zx0}
\zeta=-\fr{V_N}{V}=\fr{6x}{1+x}
\ee
and if we invert eq.(\ref{zx0}) we simply get
\be\la{xs}
x=\fr{\zeta}{6-\zeta}
 \ee
and an EoS
\be\la{eqw}
w=\fr{x-1}{x+1}=-1+ \fr{\zeta}{3}.
\ee
Using eq.(\ref{eqw}) we have $3-\zeta=-3w$ and eq.({\ref{ys1}) becomes then
\be\la{ys2}
y=y_i e^{3w(N-N_i)}= y_i\le(\fr{a}{a_i}\ri)^{3w}
\ee
which is valid for $\zeta$ constant.
We clearly see that the value of  $\zeta$ determines
the asymptotic behavior of $x,y$ and of the scalar field energy density $\rp$. If $\zeta< 3$ the scalar field
will dominate since $w<0$ and we will have $y\rightarrow 0, \Op\rightarrow 1$ and $x=\zeta/(6-\zeta)<1$, while for
$\zeta=3$ we have $x=1,w=0$ and $y$ constant given in eq.(\ref{yop}). Finally for $\zeta>3$ we get $x>1, w>0$
and $y\rightarrow \infty$ with $\Op \rightarrow 0$. Of course an accelerated universe requires $w<0, x<1$ and therefore
$\zeta <3$ close to present time. However, in general the value of $\zeta$  will be time (or $N$) dependent.

\subsubsection{ $\zeta$ as a function of $x$ and $y$}

The function  $\zeta$  is a function of $\p$ and $\p_N$, as we can see form eq.(\ref{zz}) and we
can express $\p_N$ in terms of $x,y$ as
\be\la{pn2}
\p_N = {\rm sign}[\dot \p]\; \sqrt{\fr{6 x}{1+x+y}}
\ee
(for a rolling scalar field $\dot\p>0$ but we have kept the term ${\rm sign}[\dot \p]$ for completeness),
\be\la{zz3}
\zeta=  - \fr{V_N}{V} = {\rm sign}[\dot \p]\;  \lm  \; \sqrt{\fr{6 x}{1+x+y}}
\ee
and the eqs.(\ref{dxyn}) read now
\bea\la{dxyn2b}
x_N  =&& - 6x + {\rm sign}[\dot \p]\;(1+x)\;  \lm  \; \sqrt{\fr{6 x}{1+x+y}}\\
y_N  =&&  -3y + {\rm sign}[\dot \p]\;y \;  \lm  \; \sqrt{\fr{6 x}{1+x+y}}.
\eea
The amount of $\Op$ can be determined at the critical point $x_N=y_N=0$
giving $x=1, w=0$ (i.e. the scalar field redshifts as the barotropic fluid
which in our case is matter $w_m=0$)  and
\bea\la{yop}
y &&=\le(\fr{\lm^2}{\zeta}-1\ri)\le(\fr{6}{6-\zeta}\ri)=2\le(\fr{\lm^2}{3}-1\ri),\\
\Op &&=\fr{\zeta}{\lm^2}=\fr{3}{\lm^2}.
\eea
This solution is valid for $ \lm^2 > 3$ and has a $\zeta=3$. The critical point for
 $\lm^2<3$  has
\be
 x=\fr{\lm^2}{6-\lm^2},\;\; \;\; w=-1+\fr{\lm^2}{3},\;\; \;\;  y\rightarrow 0,\;\; \;\; \Op\rightarrow 1.
\ee

\subsection{Dynamical System  for  $\hat{x}\equiv \dot \p/\sqrt{6}H$ and  $\hat{y}\equiv \sqrt{V/3H}$}\la{secxyl}

As a matter of completeness we present  here also  the dynamical evolution equations in term of the variables
$\hat{x}\equiv \dot \p/\sqrt{6}H=\p_N/\sqrt{6} $ and  $\hat{y}\equiv \sqrt{V/3H}$ which have
been widely used in the literature for studying the dynamics of scalar fields  \ci{quint.ax}. In this case the
dynamical equations are given by n
\bea\la{dxyn2}
\hat{x}_N&=& -3\hat{ x} +  \sqrt {3 \over 2} \lambda\, \hat{ y}^2 - \hat{x}\fr{H_N}{H} \non \\
\hat{y}_N&=& - \sqrt {3 \over 2} \lambda \,\hat{ x}\, \hat{y}  - \hat{y} \fr{H_N}{H}
 \la{cosmo1} \\
 \fr{H_N}{H}&=& -{3 \over 2}[1+\hat{x}^2-\hat{y}^2 ] \non
\eea
valid for a flat universe with a scalar field and matter.
These equations depend only on the function $\lambda (\p(N)) =-V'/V$.
The variables $x,y$ in eqs.(\ref{x}),(\ref{yd}) and (\ref{zz}) are related to $\hat{x},\hat{y}$ by
\be
x\equiv  \fr{\dot\p^2}{2V} =\fr{\hat{x}^2}{\hat{y}^2},\;\;\;\; y\equiv  \fr{\rmm}{V}=\fr{1-\hat{x}^2-\hat{y}^2}{\hat{y}^2},
\ee
\be
\zeta=-V_N/V=\lm \,\sqrt{6}\, \hat{x}.
\ee
we can write
\bea
\hat{x}\equiv && \fr{\dot \p}{\sqrt{6}H}={{\rm sign}}[\dot\p]\;\sqrt{\fr{x}{1+x+y}},\\
\hat{y}\equiv && \sqrt{\fr{V}{3H}}=\sqrt{\fr{1}{1+x+y}}
\eea
with   $\Op=\hat{x}^2+\hat{y}^2$. We can easily verify that $x_N=2x(\fr{\hat{x_N}}{\hat{x}}-\fr{\hat{y_N}}{\hat{y}})$
in eq.(\ref{dxy}) is equivalent to  eqs.(\ref{cosmo1}).
The critical points of eqs.(\ref{cosmo1}) have been determined in \ci{quint.ax} as a function of $\lm$
giving for $\lm^2 >3$
\be
\hat{x}=\sqrt{\fr{3}{2\lm^2}},\;\;\;\;\hat{y}=\sqrt{\fr{3}{2\lm^2}}, \;\;\;\;\Op=\fr{3}{\lm^2}
\ee
and therefore $x=\hat{x}^2/\hat{y}^2=1, w=0$  with $\zeta=\lm \,\sqrt{6}\, \hat{x}=3$ as in eq.(\ref{xs}). On the other hand
if $\lm^2 < 3$ then one finds
\be
\hat{x}=\sqrt{\fr{\lm^2}{6}},\;\;\;\;\hat{y}=\sqrt{1-\fr{\lm^2}{6}}, \;\;\;\;\Op=1
\ee
giving
\be\la{xwz}
x=\fr{\hat{x}^2}{\hat{y}^2}=\fr{\lm^2}{6-\lm^2}, \;\;w=-1+\fr{\lm^2}{3}, \;\;\zeta=\lm^2
\ee
with $\lm \,\sqrt{6}\, \hat{x}=\lm^2$ and we recover $x$ as in eq.(\ref{xs}). An universe with a late time acceleration has then $\zeta\simeq \lm^2 < 3$
which corresponds to  the slow roll approximation.

\subsection{Summary on $L$, $\zeta$ and $\lm$}\la{secLzl}

We have defined three different quantities $L$,  $\lm$ and $\zeta$   given in eqs.(\ref{lq}) and (\ref{zz}).
Both $\zeta$ and $\lm$ determine uniquely the set of differential equations
eqs.(\ref{dxyn}) and (\ref{dxyn2}),  respectively. The solution to these
equations then depend on the initial conditions $x_i, y_i$ and on the functional
form of $\zeta$  or $\lm$ as a function of $N$. These three quantities are related by eqs.(\ref{lq}) and
(\ref{zq})
\be
L=\lm^2 (1+q)^2=\le(\fr{6x\lm}{\zeta}\ri)^2,
\ee
and in the  slow roll approximation $|q|\ll 1$
they  coincide  since $L\equiv\lm^2(1+q)^2\simeq \lm^2\simeq \zeta$.
Therefore all three quantities are equivalent in the full slow roll approximation
and differ slightly once the evolution of $\p$ does not obey it any more. All of them
have advantages and can be used to determine the evolution of $\p$ uniquely.

If we determine $\lm$ and $y=\rmm/V$ as a function of the scale factor $a$, then we can
extract $\p$ as a function of the scale factor  using $dV/V= \lm \phi_N dN$  and $\p_N=\sqrt{xV}$. We get
\be\la{ddp}
V=V_i e^{\int  \lm \p_N dN},\;\;\p=\int \sqrt{x y \rmm} dN
\ee
with  $V(a)=\rmm(a)/y(a)$. From  eq.(\ref{ddp}) we   have $V(\p)$ and   $\p$ as a function of the scale factor
and we can  then  determine  $V(\p)$ as a function of $\p$.

We have studied the critical points for $x,y$ or $\hat x,\hat y$ for constant $\zeta$ or $\lm$, respectively,
which give the asymptotic behavior of the system. However, we do not expect in general to
have $\zeta$ (or $\lm$)  constant and we would need to obtain
the dynamical equation of motion for  $\zeta$ ($\lm$). This can be easily done by taking for example
the derivative $\zeta_N$, i.e. $\zeta_N=V_{NN}/V-V_N^2/V^2= \zeta^2(V_{NN}V/V_N^2-1)$, and we can
use the equation of motion of the scalar field to determine the rhs of the
equation of $\zeta_N$. This is not a closed system and we could take further derivatives
of $\zeta$ and have an infinite  series of equation in which the functions
$d^n\zeta/d^n N$ are given in terms  of  $d^m\zeta/d^m N$ with $m<n$.
In order to have  an exact solution for the scalar field we would need to solve
the equation of motion of $\p$ given a potential $V(\p)$, but then we would have an exact solution only  for the
specific potential used.

In principle the quantity $\zeta(N)=\lm\p_N$ may have
many free parameters and may be a complicated function since it depends on the potential $V,V'$ via $\lm$
and the kinetic term $\dot\p$.
Instead of using $\zeta$ as our free function
we prefer to work with $\lm$  since it is only a function  of  the potential $V(\p)$ and not of $\p_N$.
If we take the derivative of $\lm$ we get
\be\la{lmm1}
\lm_N=\le(\lm^2-\fr{V''}{V}\ri)\p_N
\ee
and we will distinguish two different cases for $\lm_N$.
If we have a rolling scalar field in such a way that $V'$
 only  vanishes at infinite value of $\p$, i.e. there is no local minimum, then
the term
\be\la{G}
\Gamma\equiv\fr{V''V}{V'^2}
\ee
can be treated in a good approximation as constant, at least for tracker
potentials with $\G>1$ \ci{tracker}, and eq.(\ref{lmm1}) can be written as
\be\la{lmm2}
\lm_N=\lm^2\le(1-\G \ri)\p_N,
\ee
and this ansatz corresponds to a wide class of potentials  \ci{tracker}-\ci{IDE.ax}.
We show in Figs.\ref{fig1}a, \ref{fig1}b  and \ref{fig1}c the evolution of $w$ for an inverse power IPL  $V=V_o\phi^{-3/2}$, an exponential $V=V_o e^{-\phi/5}$ and a sugra $V=V_o \phi^{-4} e^{\phi^2/5}$ potentials, respectively, for different initial conditions and with $\Omega_{\rm DE o}=0.7$.
Notice that for the same potential we have different evolution of $w$ and it may grow or decrease as a function of  z
and we have a steep transition which also depends on the initial conditions. We show
in figure \ref{FI}a the evolution of  for different initial conditions $\hat x_i, \hat y_i$ and same
$\G, L_i$, where $L_i$ is the initial condition of $L$. It is clear that  $\hat x, \hat y$  and the EoS $w$ have an attractor
solution but what figures \ref{FI}b,c show is that
the model may not have reached the attractor solution by the time one has $\Op=0.7$ and the evolution
of $w$ and $L$ as a function of  z  depends strongly not only on the value of $\G$ but also on the initial conditions
$\hat x_i, \hat y_i$ and $L_i$.
We have used a $\G=1.1+0.5k$ with $k=1,2,3,4$ (green,red,yellow and blue respectively and $L_i=1+10j$ with $j=0,1,3,5,7$.
\begin{figure}[h!]
    \includegraphics[width=3in]{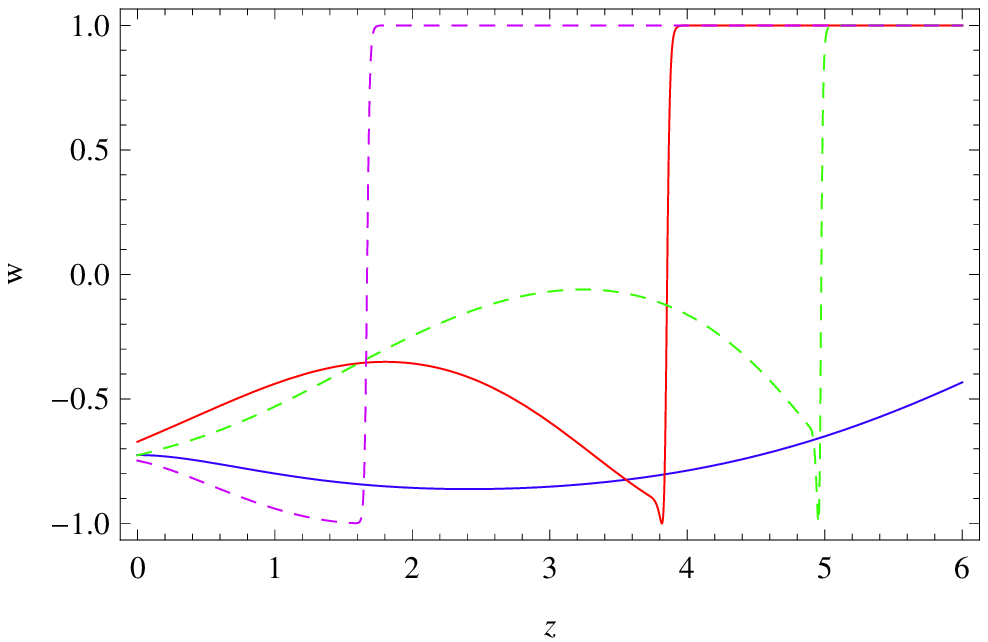}
    \includegraphics[width=3in]{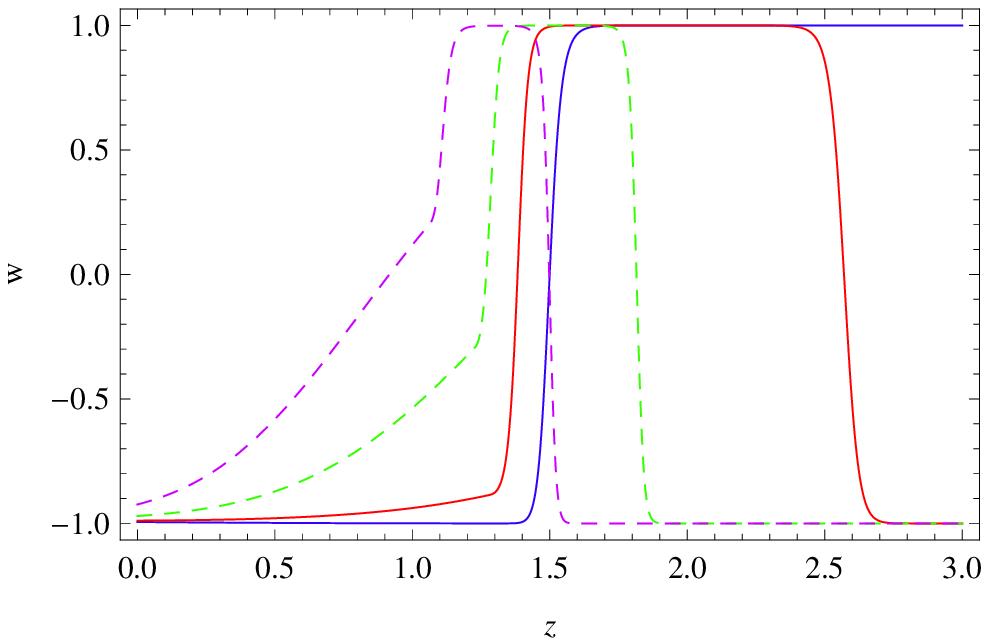}
    \includegraphics[width=3in]{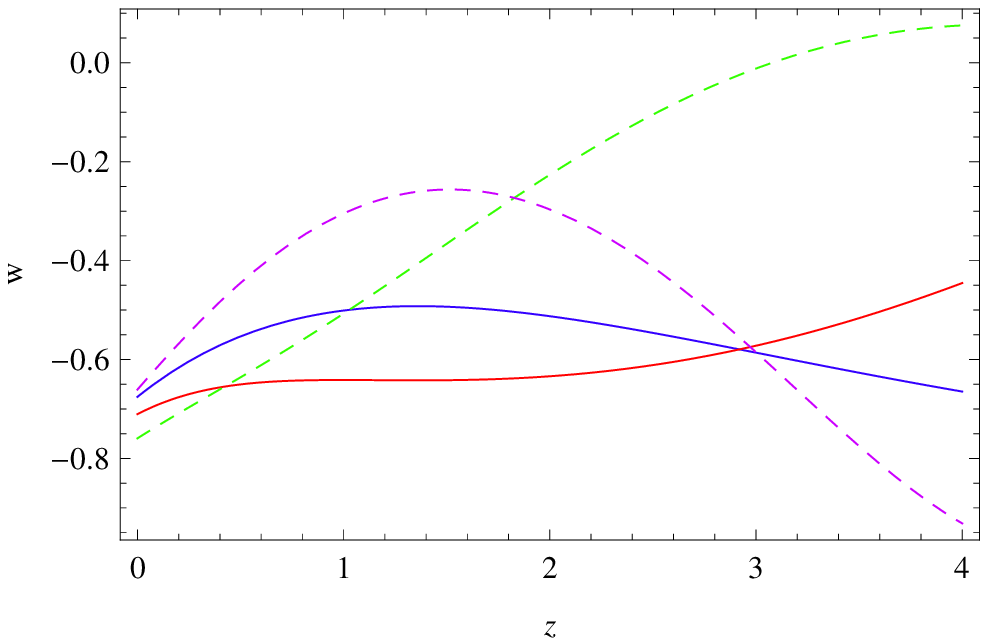}
      \caption{\footnotesize{In Figs.\ref{fig1}a, \ref{fig1}b  and \ref{fig1}c we show the evolution of $w$ for a potential $V=V_o\phi^{-3/2}$, $V=V_o e^{-\phi/5}$ and $V=V_o \phi^{-4} e^{\phi^2/5}$, respectively, for different initial conditions.}}
  \label{fig1}
\end{figure}

\begin{figure}[h!]
    \includegraphics[width=3in]{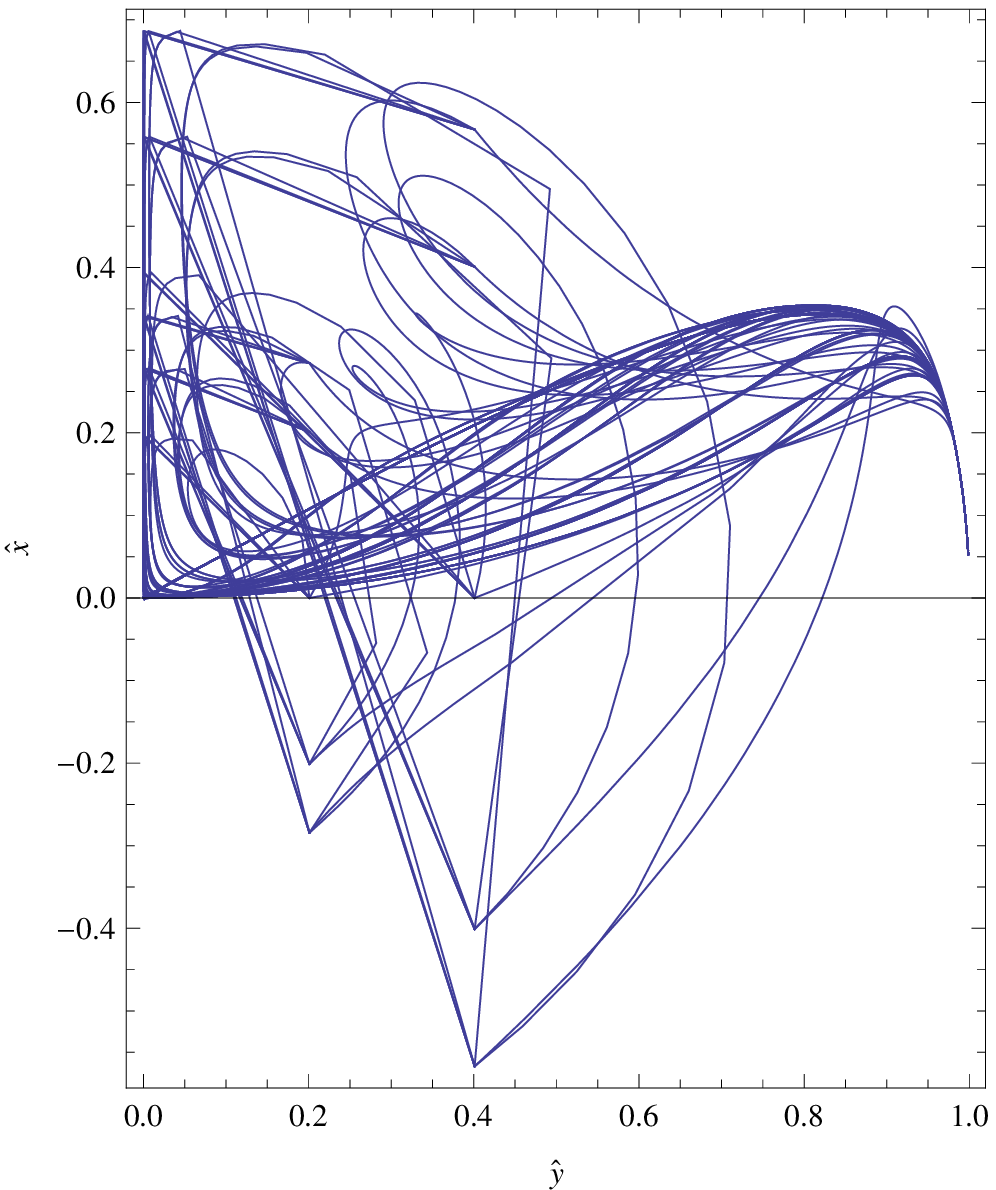}
    \includegraphics[width=3in]{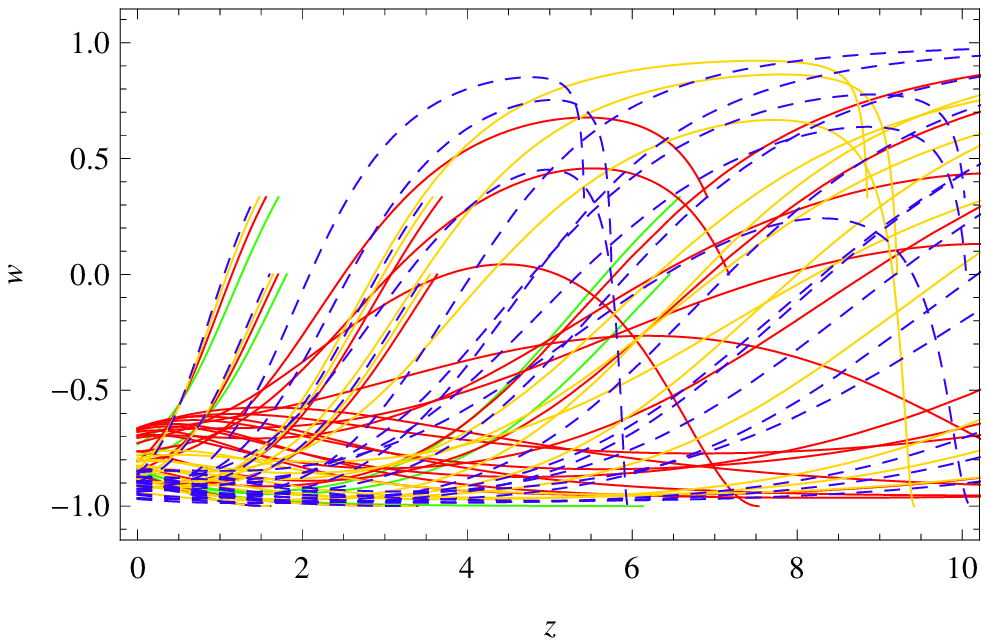}
      \includegraphics[width=3in]{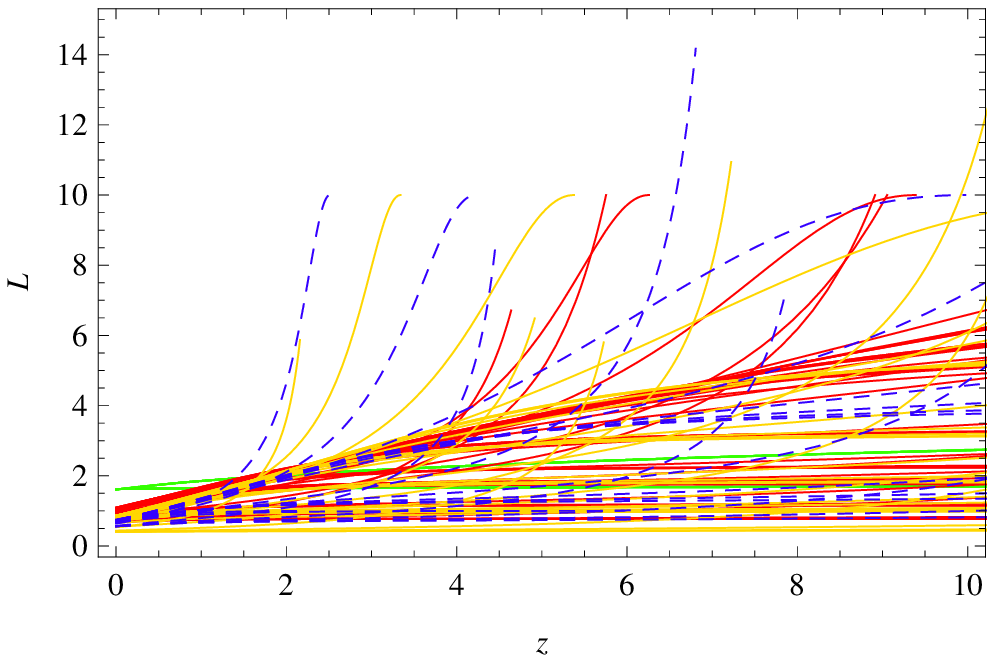}
      \caption{\footnotesize{In fig.\ref{FI}a we show the evolution of $\hat x, \hat y$ for different initial conditions. We also
      show in fig.\ref{FI}b,c $w$ and the corresponding $L$ as a function of  z  for different values of $\G$ and $L_i,\hat x_i, \hat y_i$. }}
  \label{FI}
\end{figure}

\section{Scalar Field DE parametrization}\la{sfp}

The aim here is to test a wide class of DE models in order
to constrain the dynamics  of Dark Energy form the observational data.
We could  parameterize $w$ or $x$ with a single function in terms of the
scale factor (c.f. Sec.\ref{o.p.})  but as we have seen in eq.(\ref{x3}) we get a better understanding of
the evolution of $\p$ if we parameterize  the functions $L$ and $y$.

In order to have the evolution of Dark Energy we need to either
choose a potential $V(\p)$  and solve eq.(\ref{dxyn}) or parameterize the EoS
$w$ or  $x$ as a function of the scale factor $a$. If we choose to parameterize
$\zeta$ or $\lm$ then we need to solve the differential equations  eq.(\ref{dxyn}) or  (\ref{dxyn2}).
On the other hand $L$
gives the evolution of $x$ and $w$ without needing to solve  any differential equations
and $L$ reduces to  $\lm$ (or  $\zeta$) in the slow roll approximation.
Therefore we propose here to use the exact equation for $w$  given in eq.(\ref{www})
and parameterize the function $L$ and $y$.

A priori it is impossible to know how many free parameters
has  the potential $V(\p)$ and since the evolution of different scalar field model
requires the solve the  eq.(\ref{dxyn}) and one has a difficult task to test a wide range
of potentials $V$. The potential $V$ may involve  a single term, as in a runaway  potential  as
in  $V=V_o\phi^{-n}$ or $V=V_o e^{-n \phi}$, or it
may have different terms in the potential  with the same order of magnitude as for example $V=A\phi^{-1}+B  e^{n \phi}$ or $V=A\phi^{-1}+ m^2 \phi^2$ that may lead to  a local minimum with $V\neq 0$.

In all these cases the evolution of $w$  would be  different but still constrained to  $-1\leq w \leq 1$, and
in the slow roll approximation ($ L\ll 1, x\ll 1$) one has  $w\simeq -1 +L/[3(1+y)]$ given by eq.(\ref{lm1}).
However, we do not know if there was a steep descent in
$w$ or $x$ close to present time in such a way that for smaller $a<a_o$ the value of $w$ was
much larger, e.g. $w$ close to 0, as matter,  or even positive  $w\simeq 1/3$, as radiation,
with  $L(a<a_o) \gg L(a_o)$ and the slow roll approximation may not be valid anymore.
In this work, we would like to include a parametrization which includes also this kind
of steep transition.

How many parameters should we use? We would say that we
should use the minimum number of parameters as long as we still track the
behavior of scalar fields and is generic enough to have different behavior
for $w$ and allow the cosmological data to fix these parameters.

\subsection{Ansatz}\la{ans}

We will propose an anzatz for $L$ and $y$ that covers
the generic behavior of scalar field  leading to an accelerating universe.
If we want to have a constant EoS for DE  at early times $z\gg 1$, as for example
matter $w=0$ or radiation $w=1/3$, which
are reasonable behavior for particles, we should choose $L$ proportional to $y$ for large $y$
with $L/y \rightarrow L_1$,
or if we want  $w\rightarrow -1$ at a large redshift the  limit $L/y \rightarrow 0$ must be satisfied.
We  then propose to take
\be\la{ll}
L=L_o+L_1\, y^\xi \,f(z)
\ee
with two  free constant parameters $L_o$ and $L_1$, a transition function $f(z)$ constrained between  $0\leq f(z)\leq 1$.
The quantity  $\xi$ takes the values $\xi=1$ and $\xi=0$ only and we do not consider it as a free parameter but more as two different ansatze for $L$, depending if we want $L$ to grow proportional to $y$ (i.e. $\xi=1$) or to a constant value in which case we take $\xi=0$.
Since in the limit of large redshift  $z\gg 1$ we have  $y\gg1$ and $L/y $ goes  to a constant value
$L_1$ or zero for $\xi=1,0$ respectively. This limit is independent of the functional form of $y$,
since at large $y$ the EoS $w$ depends on $L/y\rightarrow L_1$ and the dependence on $y$ cancels out giving a constant value and a constant $w$ (c.f. eq.(\ref{lm3})). We therefore choose to take $y$ in eq.(\ref{ll}) as in eq.(\ref{ys2}) with $w\simeq w_o$, i.e.
\be\la{yy}
y=\fr{\rmm}{V}=  y_o \le(\fr{a}{a_o}\ri)^{3w_o}= y_o (1+z)^{-3w_o}.
\ee
However, taking $y$ as in eq.(\ref{ys2})  does \emph{not} mean that
$\rp\propto  (a_o/a)^{3(1+w_o)}$ since the kinetic energy $\dot\p^2/2$, or equivalently $x$,  may grow faster or slower than $V$.
For $\xi=1$,  eq.(\ref{ll}) allows a wide class of  behaviors for $w$.  If we want $w$ to increase
to $w=0,1/3$ at earlier times we would take $L_1=6,12$, respectively, or since  in many scalar field models the evolution of $w$ goes form $w_o$ to a region dominated by the kinetic energy density with $w=1$   and in this case we would
should take $L_1 \gg 1$. Of course a
$w(z\gg 1)=1 $ would only be  valid for a limited period  since $\Op$ should not dominate the universe at early times.
We have included in eq.(\ref{ll}) the case $\xi=0$ because we want to allow $w$ to  increase from $w_o$ at small
 z  and later go to $w\rightarrow -1$ (c.f. pink-dashed line in fig.(\ref{FC})), since this is the behavior of potentials used as a models of DE as for example  $V=V_o\phi^{-n},n=2/3$ derived from gauge group dynamics \ci{GDE.ax} where the behavior
of $w(z)$ close to present time depends on the initial conditions.

We propose to take the simple ansatz for the transition $f(z)$ given by
\be\la{f2}
f(z)=\fr{(z/zt)^k}{1+(z/z_t)^k},
\ee
with $f_o\equiv f(z=0)=0, f(z=z_t)=1/2$ and $f_i\equiv f(z\gg 1)=1$. The transition function $f(z)$ given in eq.(\ref{f2}) has four free parameters $L_o, L_1, z_t$ and $k$.
The parameters $L_o$ and $L_1$ give the EoS $w$ at present time and at large redshift $z \gg z_t$,
while the transition epoch from $w_o$ to $w_i\equiv w(z\gg z_t)$ is given by $z_t$ and $k$ sets the steepness of this transition.
This parametrization has a simple expression
in terms of  z  and it reduces to CPL \ci{CPL}, i.e. $w=w_o+w_a(1-a)$),  in the  full slow roll approximation
for  $\xi=0$  and $z_t=1, k=1$ one has $f(z)=z/(1+z)=1-a$ and $L=L_1+L_o(1-a)$. Using eq.(\ref{at2}) we have
$w\simeq-1+L/3=-1+(L_o+L_1f(z))/3$ and therefore we identify  $w_o=-1+L_o/3$ and $w_a=L_1/3$.
However, our parametrization given in eqs.(\ref{ll}) and (\ref{f2}) goes well beyond the CPL parametrization.

\begin{figure}[h!]
    \includegraphics[width=3in]{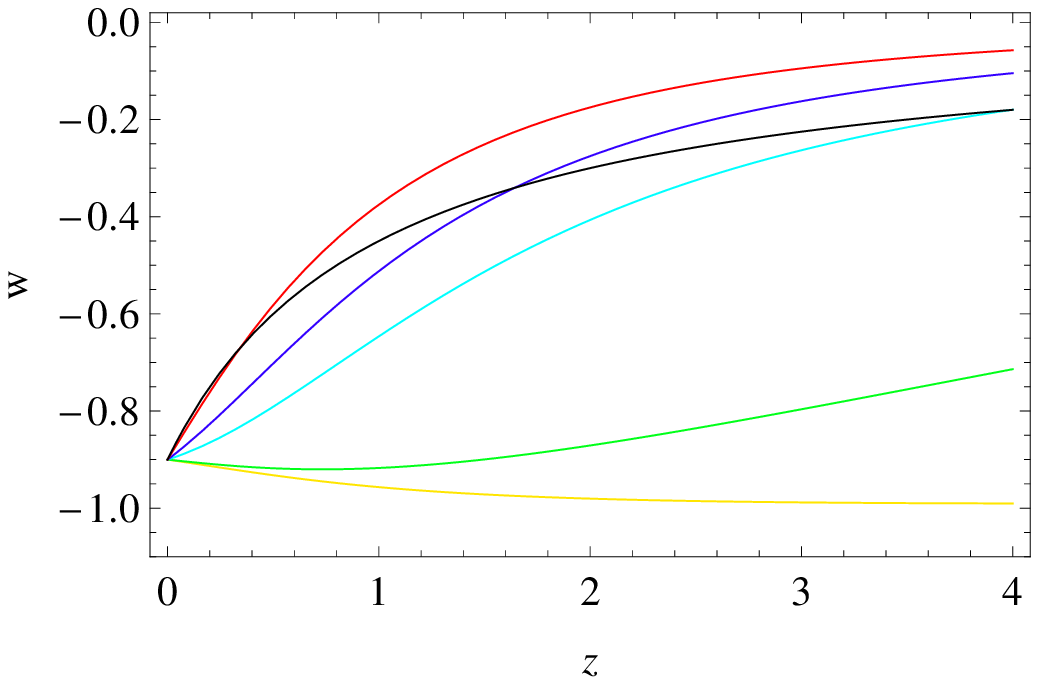}
    \includegraphics[width=3in]{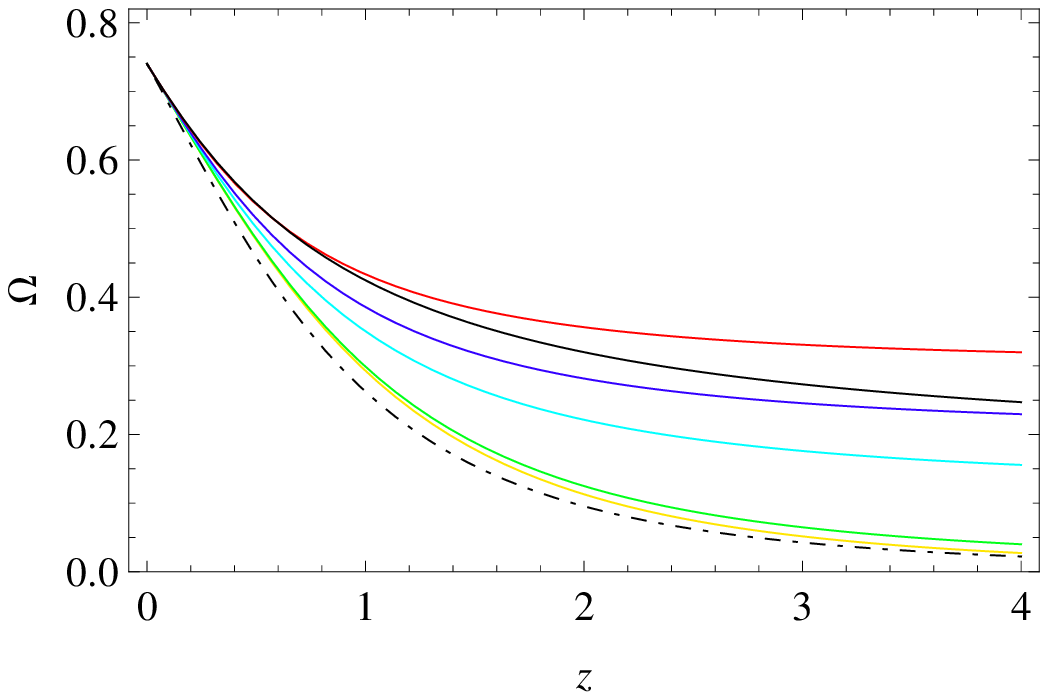}
      \includegraphics[width=3in]{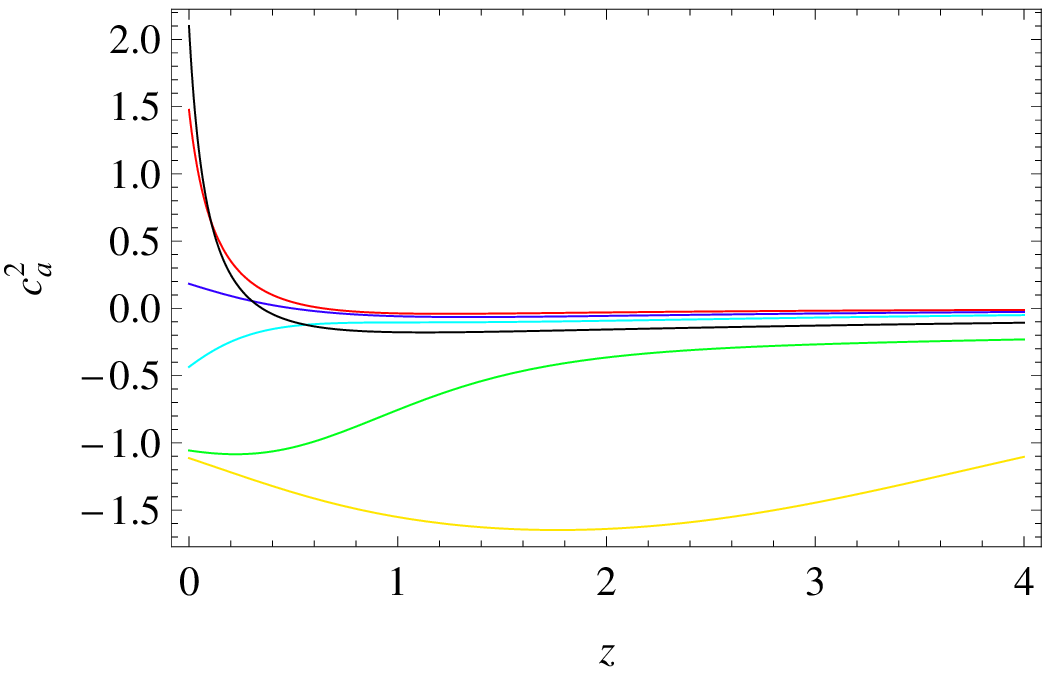}
      \caption{\footnotesize{We show the evolution of  $\Op$, $w$ and $c_a^2$ for different models.  We have taken
    $\xi=1$, $k=2$,$ L_1=6$ with $z_t=0.1,1,$$2,10,100$ (red, dark blue, light blue, green and yellow, respectively).
   In black we have $w,\Omega_w$ using $w$ in eq.(\ref{w1}) and $\Omega_{cc}$ for a cosmological constant (black dot-dashed).  We take in all cases $w_o=-0.9$, $\Opo=0.74$ and $L_1=6$ giving $w_i=w(z\gg z_t)=0$ for large  z .}}
  \label{FA}
\end{figure}
 \begin{figure}[h!]
 \includegraphics[width=3in]{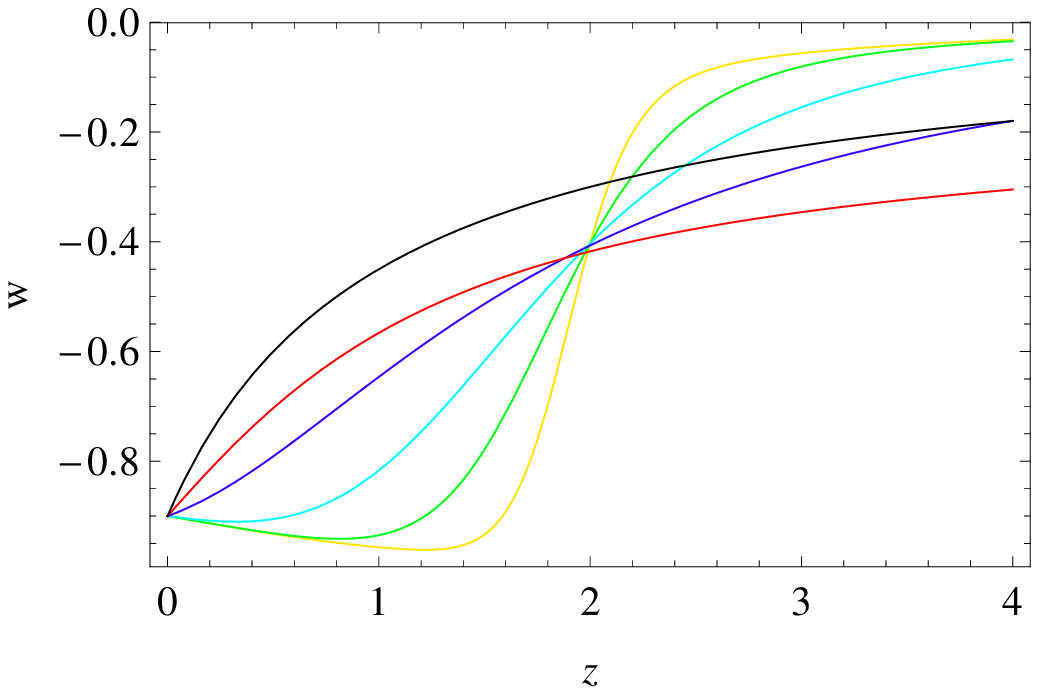}
  \includegraphics[width=3in]{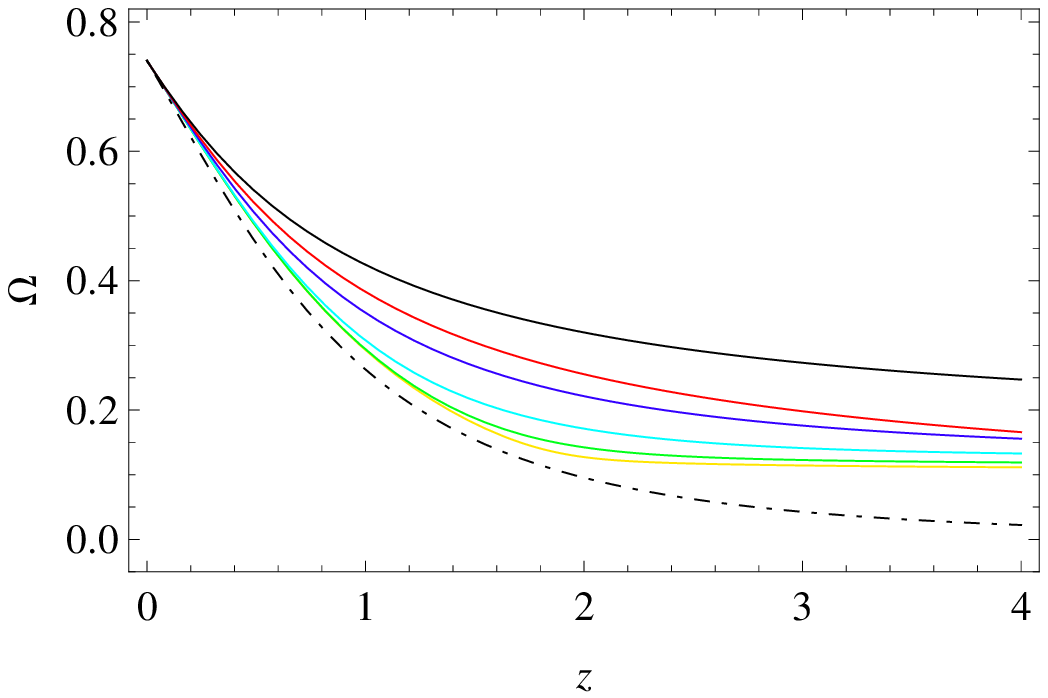}
    \includegraphics[width=3in]{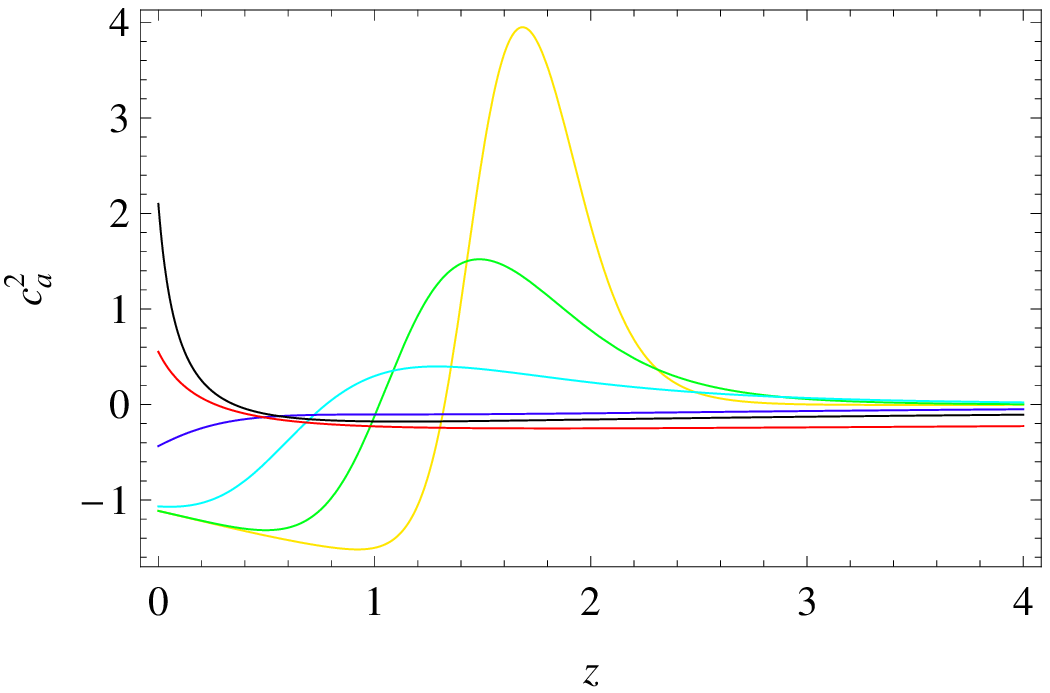}
      \caption{\footnotesize{We show the evolution of  $\Op$, $w$ and $c_a^2$ for different models.  We have taken
    $\xi=1$, $ z_t=2$,$L_1=6$ fixed and $k=1/2,$ $2,$ $5,10,20$ (red, dark blue, light blue, green and  yellow, respectively)
   In black we have $w,\Omega_w$ using $w$ in eq.(\ref{w1}) and $\Omega_{cc}$ for a cosmological constant (black dot-dashed).
   We take in all cases $w_o=-0.9$, $\Opo=0.74$ and $L_1=6$ giving $w_i=w(z\gg z_t) =0$ for large  z . }}
  \label{FB}
\end{figure}
 \begin{figure}[h!]
 \includegraphics[width=3in]{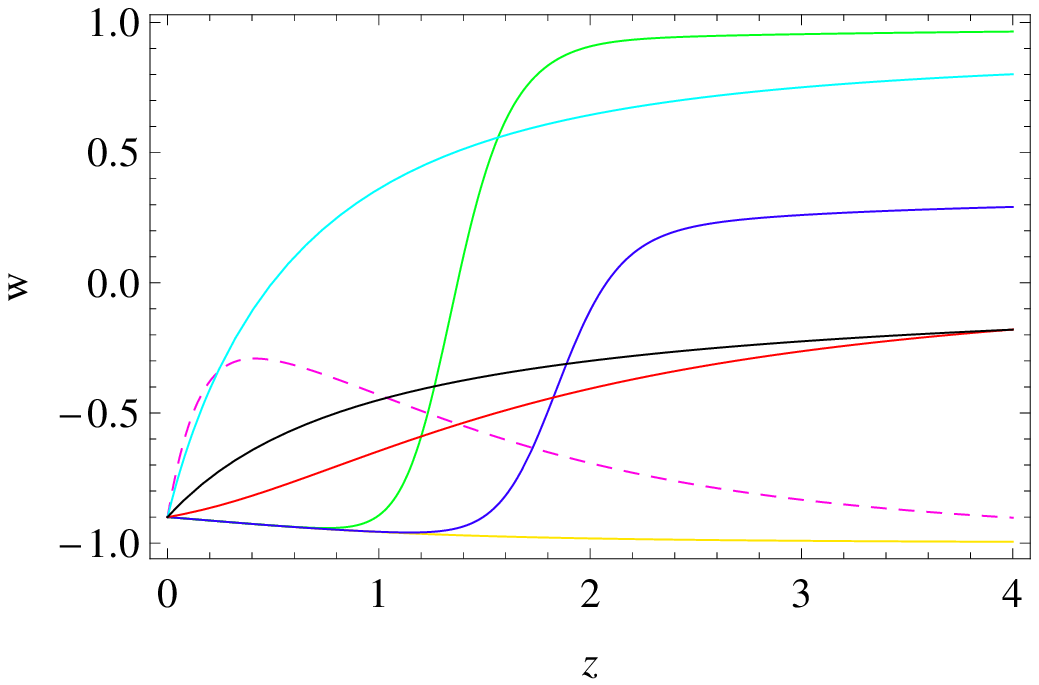}
  \includegraphics[width=3in]{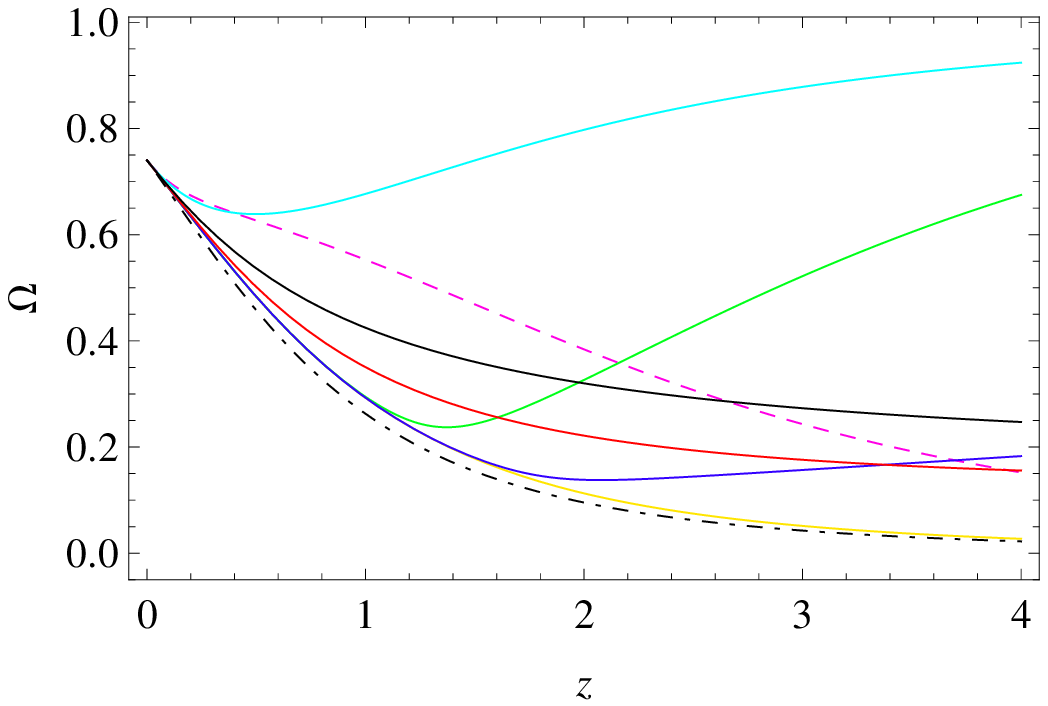}
    \includegraphics[width=3in]{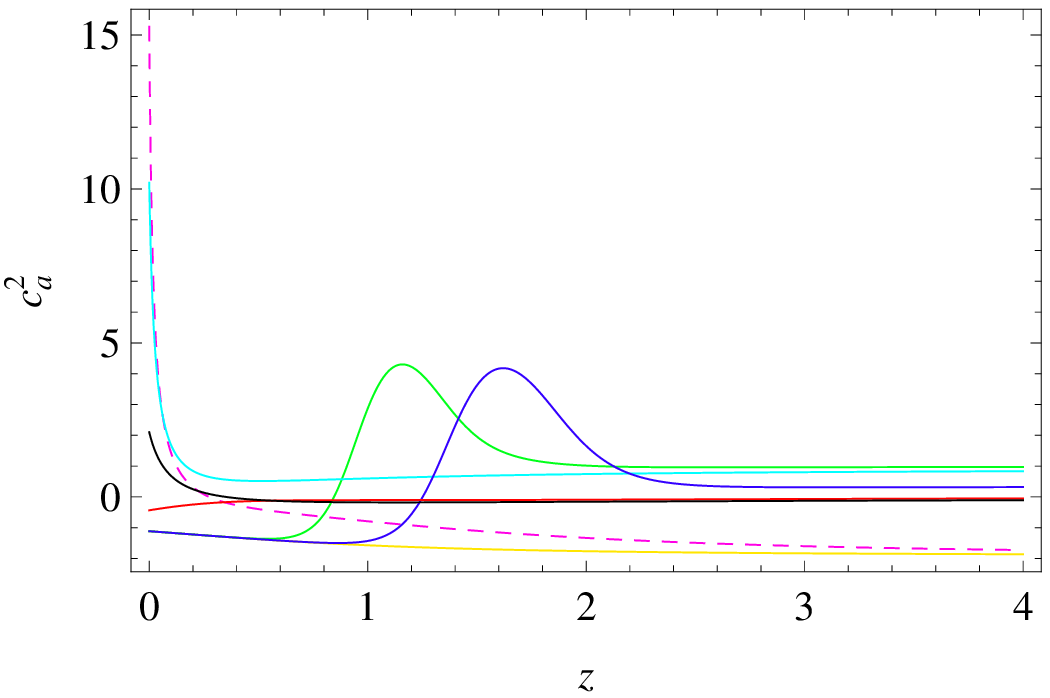}
      \caption{\footnotesize{We show the evolution of  $\Op$, $w$ and $c_a^2$ for different models.  We have taken
    $\xi=1$, $z_t=2$ fixed and $(k,L_1)=(2,6),$$(20,12),(2,100),(20,1000),(20,0)$ (red, dark blue, light blue, green and  yellow, respectively) and $\xi=0$ with  $z_t=0.1, k=10, L1=12$ in pink-dashed line.
   In black we have $w,\Omega_w$ using $w$ in eq.(\ref{w1}) and $\Omega_{cc}$ for a cosmological constant (black dot-dashed).
   We take in all cases $w_o=-0.9$, $\Opo=0.74$ and $L_1=6$ giving $w_i=w(z\gg z_t) =0$ for large  z . }}
  \label{FC}
\end{figure}

\subsection{Initial Conditions and Free Parameters}

Let us summarize the parameters and initial conditions of our parametrization. The EoS $w$ is only a function of $x$ and
$x$ is a function of $L$ and $y$. From eq.(\ref{yw}) we see that $y$ depends on two parameters $w_o$ and $\Opo$
(or equivalently on $y_o,w_o$), since we are assuming a flat universe with DE and matter and  $\Omo=1-\Opo$.

From eqs.(\ref{ll})  we have the conditions at present time as
\be
y_o\equiv y(z=0) =\fr{\rmmo}{V_o}=\fr{2}{(1-w_o)}\fr{\Omo}{\Opo}
\ee
and from eq.(\ref{yw}) we express $L_o$ as a function of $w_o$ and $y_o$ we have and
\be\la{lo}
L(z=0)=L_o = \fr{12(1 + w_o) + 6 y_o(1 - w_o^2 )}{(1 - w_o)^2}
\ee
since $f(z=0)=0$. While the value at an early time we get
\be
y_i\equiv y(z\gg 1)= \fr{2}{(1-w_i)} \fr{\Omega_{mi}}{\Omega_{\phi i}}
\ee
and
\be\la{li}
L_i\equiv L(z\gg 1)=L_o + L_1 y_i^\xi f_i = \fr{12(1 + w_i) + 6 y_i(1 - w_i^2 )}{(1 - w_i)^2}\simeq \fr{6 y_i(1 + w_i )}{(1 - w_i)},
\ee
with $f_i=f(z\gg 1)=1$. Since  we expect to have negligible DE at early times we have $y_i\gg 1$ and we keep in the r.h.s. of eq.(\ref{li})  only the term proportional to  $y_i$. We see  from this equation that
for  $\xi=1$  the value of $L_1$ is determined by the early time EoS $w_i$ with  (c.f. eq.(\ref{lm3}),
\be
L_1 =\fr{6(1+w_i)}{1-w_i}
\ee
giving for example  $L_1=6$ for $w_i=0$,  $L_1=12$ for $w_i=1/3$ while  $L_1=0$ requires $w_i\rightarrow -1$ and $L_1\gg 1$  has $w_i\rightarrow 1$. However, when $\xi=0$ we have the limit $L/y\rightarrow 0$ with  $L_i=L_o + L_1$  a finite constant so  eq.(\ref{li}) requires $w_i=-1$ independently of the value of $L_i$.

From  eq.(\ref{ll}) we have that $L$ depends on  $y$ and $L_o, L_1,z_t,k$ and $w_o,\Opo$.
However, not all parameters are independent, since $L_o$ is a function of $w_o,\Opo,L_1,z_t,q$ and we are left with $\Opo$ and four parameters in $w$. To conclude, we can take the free parameters in the EoS as  $w_o$, $w_i$, the transition redshift $z_t$
and the steepness of the transition $k$.

\subsection{Other Parameterizations}\la{o.p.}

We present here some widely used parameterizations and we compare them with our DE parametrizations in this  work.
We begin with a simple  DE parametrization  \ci{CPL} given in terms of only two parameters and widely used in most
data analysis projects:
\be\la{w1}
w(a)=w_o+w_a(1-a)=w_o+w_a\fr{z}{1+z}
\ee
with  the derivative
\be
\fr{dw}{da}=-w_a, \;\;\;\; \fr{dw}{dz}=\fr{da}{dz}\fr{dw}{da}=(1+z)^{-2}w_a
\ee
Clearly $w$ in eq.(\ref{w1}) is convenient since it is a simple EoS and it has only two parameters.
However, it may be too restrictive  and we do not see a clear connection between the value of
$w$ at small $a$ and its derivative at present time $w_a$. It has only 3 parameters $\Opo,w_o,w_a$,
two less than our model but our model has a much richer structure.

Another interesting parametrization was presented in \ci{corasaniti}, with  4 free parameters. It is given by
 \be\la{wg}
 w=w_o+(w_i-w_o)\,G,\;\;G\equiv \fr{1+e^{a_{d}/d}}{1-e^{1/d}}\fr{1-e^{(1-a)/d_i}}{1+e^{(a_{d}-a)1/d}}
 \ee
 where $w_o,w_i,a_d, d $ are constant parameters. The function $G$ is constraint between $0\leq G\leq 1$ with
 $G=0$ for $a\gg a_{d}$ and $G=1$ for $a\ll a_{d}$. Therefore
 $a_{d}$ is the scale factor where the transition of the EoS  $w$ goes from $w_o$ to $w_i$
The parameter $d$ gives the width between the transition, for small $d$ the transition between $w_o$ and $w_{1}$ is steep.  Even though w in eq.(\ref{wg}) gives a large variety of DE behavior \ci{corasaniti}, since the sign of the slope is fixed our parametrization in eqs.(\ref{x3}) and (\ref{ll}) has a richer structure with the same number of parameters.

In the work of \ci{huang} they have followed a similar motivation  as in the present work.
They have presented two  DE parametrizations motivated by the dynamics of a scalar field.
Their parametrizations  have either two or three free parameters, and are given in eqs.(25) and (28) in the paper \ci{huang},
respectively (they do not take $a_{eq}$ as a free parameter but we do think that it is an extra parameter).
The two parameters involved determine  the quantities $w_i(a\gg a_o)$, which gives the
EoS at an early time, and $\lm(a_{eq})=V'/V|_{a_{eq}}$  at DM-DE equality (i.e. $\Om=\Omega_{de}$). In the second
case,  the parametrization also involves  a term $\zeta_s$ (eq.(23) in \ci{huang}) which depends
on a second derivative of $V$ and on the value of $\dot\phi/H$ at DM-DE equality.  Since the functional form of
the evolution of the EoS $w(a/a_{eq})$  is fixed in their parametrization the value of $w_o$ at present time is determined if we know  the value of $a_o/a_{eq}$. Therefore, the quantity $a_{eq}$ must also be assumed as a free parameter.   As in our present work,
the system of equations do not close without the knowledge of the complete $V$ as a function of $\p$. However,
since we are both interested in extracting  information from the observational data to determine the scalar
potential the parametrization, given in eqs.(25) and (28) in \ci{huang}, is an interesting  proposal to study a wide range
of potentials $V$. Here we have taken a different   parametrization which has a closer connection to
the scalar potential $V(\p)$ given by  eqs.(\ref{x}) and (\ref{w}).

\subsection{Results}

We have plotted $w$ for different sets of the parameters in figs.(\ref{FA}),(\ref{FB}) and (\ref{FC})
to show how $w$ depends on $L_o$, $L_1$, $z_t$ and $k$.  We notice that our parametrization in eq.(\ref{ll})
has a very rich structure allowing for $w$ to grow and or decrease at different redshifts. We
have also plotted $\Op$ and the adiabatic sound speed $c_a^2$ defined in eq.(\ref{ca}) for each model.
We are showing some extreme cases which do not expect to be observational valid but we want to show
the full extend of our parametrization.  We have also included a cosmological constant "C.C." (black dot-dashed) in
the figures of $\Op$ and since $w_{cc}\equiv -1$ and $c_{a\, cc}^2=0$ we do not include them in the  graphs for $w,c_a^2$.
We also plotted  $w,\Op, c_a^2$ (in black) for the parametrization in eq.(\ref{w1}) for comparison.
We take in all cases $w_o=-0.9$, $\Opo=0.74$.

In fig.(\ref{FA}) we show the evolution of  $\Op$, $w$ and $c_a^2$ for different models.
We have taken $\xi=1, k=2,\, L_1=6$ with $z_t=0.1,1,2,10,100$ (red, dark blue, light blue, green and yellow, respectively).    We take in all cases  $L_1=6$ giving $w_i=w(z\gg z_t)=0$. Notice
that  the yellow line the increase to $w=0$  does not show in the graph since $z_t=100$ and we plot $w$ only up  to
$z=4$. The slope in $w$ depends on the value of $z_t$ and since $k=2$ is not large the transition  is not  steep.
The value of $\Op$ decreases slower than for a C.C. and for smaller $z_t$ it decreases even more slowly, i.e when $w$ approaches zero faster.

In fig.(\ref{FB}) we show  the evolution of  $\Op$, $w$ and $c_a^2$ for different models. In this
case we have taken $\xi=1, z_t=2,\,L_1=6$ fixed  with $w_i=0$ and we vary $k=1/2,2,5,10,20$ (red, dark blue, light blue, green and  yellow, respectively).
We clearly see in the evolution  $w$ how the steepness of the transition depends $k$ and that $c_a^2$
has a bump at $z_t$ and it is more prominent  for steeper transition. This is generic behavior and we could expect
to see a signature of the transition in large scale structure as discussed in Sec.\ref{sec.pert}.

In fig.(\ref{FC}) we show   the evolution of  $\Op$, $w$ and $c_a^2$ and we take
$\xi=1, z_t=2$ fixed and $(k,L_1)=(2,6),(20,12),(2,100),(20,1000),(20,0)$ (red, dark blue, light blue, green and  yellow, respectively)
and $\xi=0$ with  $z_t=0.1, q=10, L1=12$  (pink-dashed line).
In this case  we vary $L_1$  and  we see that for large $L_1$ the EoS $w$
becomes bigger and it may approach $w\simeq 1$ (e.g. green line). Of course this case is not phenomenologically
viable but we plot it to show the distinctive cases of our $w$ parametrization. Once again, a steep transition gives
a bump in  $c_a^2$. The pink-dashed line shows how $w$ can increase at low  z  and than approach $w=-1$.

We have seen that  our parametrization gives a wide class of $w$ behavior, with increasing and decreasing $w$.
From the observational data we should be  able to fix the parameters of  $L$ in eq.(\ref{ll}) and  we could then
have a much better understanding on  the underlying potential $V(\p)$.

\section{Perturbations}\la{sec.pert}

Besides the evolution of the homogenous part of Dark Energy $\p(t)$, its
perturbations $\delta\p(t,x)$ are also an essential ingredient in determining the nature of DE.
We work  is the synchronous gauge and the  linear perturbations with a  line element
$ds^2=a^2(-d\tau^2+ (\delta_{ij}+ h_{i j})dx^idx^j$, where  $h$ is the trace of the metric perturbations
\ci{ma,hu}. In this sect.(\ref{sec.pert}) a dot represents derivative with respect to conformal time $\tau$
and $\textsf{H}=\dot a/a=(da/d\tau)/a$ is the Hubble constant w.r.t. $\tau$, while $H=(da/dt)/a$

\subsection{Scalar Field Perturbations}

For a DE given in terms of a scalar field, the evolution requires the knowledge of $V$ and
$V'$ while the evolution of  $\delta\p(t,x)$ depends also on  $V''$ through \ci{ma,hu}
\be\la{dp}
\delta \ddot\p + 2\textsf{H} \delta\dot\p+[k^2+a^2V'']\delta\p=-\fr{1}{2}\dot h \dot\p.
\ee
Eq.(\ref{dp}) can be expressed as a function of $a$ with $\dot Y =a\textsf{H} Y_a$ for $Y=\delta\p,\p,\textsf{H}, h$ and  the subscript $a$ means derivative w.r.t. $a$  (i.e. $Y_a\equiv dY/da$), giving
\be\la{dp2}
\delta \p_{aa}+\le(\fr{3}{a}+\fr{\textsf{H}_a}{\textsf{H}}\ri)\delta\p_a +\le[\fr{k^2}{a^2\textsf{H}^2}+\fr{V''}{\textsf{H}^2}\ri]\delta\p=-\fr{1}{2}h_a \p_a.
\ee
In the slow roll approximation we have
\be\la{sl2}
\le|\fr{V''}{3H^2}\ri|=\G x < 3
\ee
where we have used eq.(\ref{x3}) and $\G\equiv\fr{V''V}{V'^2}$ given in eq.(\ref{G}).

Eq.(\ref{sl2}) implies that an EoS of DE  between $-1\leq w \leq -1/3, 0, 1/3 $, with $0\leq x\leq  1/2,1,2$, requires $\G<3/x=6, 3,3/2$,  respectively.  For a scalar field $\p$ to be in the tracking regime one requires $\G$ to be approximately  constant with $\G>1$ \ci{tracker}. Therefore the regime
$1<\G<3/x$ allows a tracking behavior  satisfying also the slow roll approximation. Here we are more interested in the late time evolution of DE and the tracking regime is not required and in fact we expect deviations from it. However, if $\G$ is nearly constant the evolution of the perturbations in (\ref{dp2}) are then given only in terms of $x$  and we can use our DE parametrization in eqs.(\ref{ll}) and (\ref{f2}) to calculate them.

We can express the slow roll parameter $\epsilon,  \Upsilon$ in terms of $\G$ and $L$ in the limit $q\ll 1$ as
\be
\epsilon \equiv \fr{1}{2}  \le(\fr{V'}{V}\ri)^2 =\fr{\lm^2}{2}, \;\;\;\;\;\;  \Upsilon \equiv \fr{V''}{V}=\G\lm^2
\ee
We have decided to use $L,\Upsilon$ instead of $\epsilon,\eta$ not to confuse the reader with the inflation
parameters and the DE ones.

\subsection{Fluid Perturbations}

The evolution of the energy density perturbation $\delta = \delta\r/\r$, $\theta=k(1+w)v$, with  $v$ is the  velocity perturbation, are
\ci{ma,hu,mukhanov,bean}
\bea\la{p.d.}
 \dot \delta =-(1+w)&&\le(k^2+9\textsf{H}^2\,[c^2_s-c_a^2]\ri)\fr{\theta}{k^2}-\fr{\dot h}{2}\nonumber\\ &&-3\textsf{H}(c^2_s-w)\fr{\delta}{1+w}
 \eea
 \be\la{p.t.}
 \dot\theta=-\textsf{H}(1-3c^2_s)\theta+ c^2_sk^2\fr{\delta}{1+w},
 \ee
and we do not consider an anisotropic stress.
The evolution of the  perturbations depends on three quantities \ci{hu, bean}
\bea\la{cw}
w&=&\fr{p}{\r} \\
\la{ca}c^2_{a}&=&\fr{\dot p}{\dot\r}=w+ \dot w \fr{\rho}{\dot\r}\\
&=&  w - \fr{\dot w }{3\textsf{H}(1+w)}= w+\fr{x_z w_x}{3a(1+w)}\nonumber\\
\la{cs} c^2_{s}&=&\fr{\delta p}{\delta \r}
\eea
where $w$ is the EoS,  $\textsf{H}$ the Hubble constant in conformal time, $c^2_{a}$  is the adiabatic sound speed and
$c^2_{s}$ is the sound speed in the rest frame of the fluid \ci{mukhanov, hu}.
For a perfect fluid one has  $c^2_{s}=c^2_{a}$ but scalar fields are not perfect fluids.
The entropy perturbation  $G_i$ for a fluid $\r_i$ with $\delta_i=\delta \rho_i/\r_i$ are
\be
w_iG_i\equiv (c^2_{si}-c^2_{ai})\delta_i=\fr{\dot p_i}{\dot \r_i}(\fr{\delta p_i}{\dot p_i}-\fr{\delta \r_i}{\dot \r_i})
\ee
where the quantities $G_i$ and $c^2_{ai}$ are scale independent and gauge invariant but $c^2_{si}$
can be neither \ci{ma, bean}. In its rest frame  a scalar filed $\p$ with a canonical kinetic term one
has $c^2_{s}=\delta p/\delta \r=1$ \ci{mukhanov, hu}.  One can relate the rest frame
$\hat{\delta},\hat{\theta}$ to an  arbitrary frame $\delta,\theta$ by \ci{bean}
\be
\hat{\delta}=\delta + 3 \textsf{H}(1+w)\fr{\theta}{k^2}
\ee
and
\be
\delta p=\hat{c}^2_{s}\delta\r+(\hat{c}^2_{s}-c^2_{a})3\textsf{H} (1+w)\r_i\fr{\theta}{k^2}.
\ee
As we see from eqs.(\ref{p.d.})-(\ref{p.t.}) the evolution of $\delta$ depends on $c_a^2,c^2_s$ and $w$.
Using eq.(\ref{cw}) and since $w$ is a function of $x(a)$   we have
\be\la{caz}
c^2_a =  w+\fr{x_z w_x}{3a(1+w)}
\ee
with $w_x/(1+w)= 2/[x(1+x)]$. From eqs.(\ref{www}) and (\ref{x3})  we  can
express  $c_a^2$ as a function of the parameters of $x$.

DE perturbations are important in distinguishing different DE models \ci{hu}-\ci{DE.p}  and at the epoch where the universe is dominated by DM and DE the total perturbation is $\delta_T = \delta\r_T/\r_T= \Omega_{\rm DM} \delta\r_{DM} + \Omega_{\rm DE}  \delta\r_{DE}$. If  $\Omega_{\rm DE}$ is not much smaller than $\Omega_{\rm DM}$, i.e. at an epoch with low redshift  z ,  then the perturbations of DE have an important contribution to $\delta_T $. Since for a scalar field $c^2_a\neq w$ and $c^2_s=1$ \ci{mukhanov, hu} we see from eqs.(\ref{p.d.}) and (\ref{p.t.}) that a bump in   $c^2_a$ may  give a significant contribution to the  evolution of  $\delta\r_{DE}$ depending  on the value of $\Omega_{\rm DE}$ (i.e. the redshift of the transition)  and the steepness of the bump \ci{DE.p.sf,DE.p}.

Finally, we can relate $V''$ in terms of the adiabatic sound speed $c^2_a$ in eq.(\ref{ca})  and its time derivative,
using $c^2_a=\dot p/\dot\r=1+2V'/3H(d\p/dt)$, giving
\be\la{cp}
\fr{dc^2_a}{dt}=\le(c^2_a-1\ri)\le(\fr{V''}{V'}-\fr{3H}{2}\le(\fr{2d\dot H}{3H^2}-(c^2_a+1)\ri)\ri)
\ee
and for  $c^2_a\neq 1$ we can invert eq.(\ref{cp}) to give
\be\la{vv}
\fr{V''}{V'}=\fr{\G\,  V'}{V}= \fr{1}{(c^2_a-1)}\fr{dc^2_a}{dt}-\fr{3H}{2}\le(w_T+ c^2_a + 2\ri),
\ee
where we have used $\dot H=dH/dt=-(\r_T+p_T)/2=-3H^2(1+w_T)/2$ with $\r_T,p_T,w_T$ the total energy density,
pressure and EoS, respectively.  In our case we have $\r_T=\rmm+\rp,\, p_T=p_m+p_\p=p_\p$ and using
$\rp=V(1+x)$ and eqs.(\ref{w}) and (\ref{h}) we have
\be\la{wT}
w_T\equiv \fr{p_T}{\r_T}= w\Op= \fr{w \rp}{3H^2}=\fr{w(1+x)}{1+x+y}=\fr{x-1}{1+x+y}.
\ee
With eq.(\ref{wT}) the l.h.s. of eq.(\ref{vv}) depends
then only on $y,x$ and are fully determined by our parametrization. In the full slow roll approximation $\ddot\p=0$
and one has $c_a^2=-1$.

\section{Conclusions}\la{con}

We have presented a new parametrization of Dark Energy  motivated by the dynamics of a canonical normalized
scalar field minimally coupled to gravity.   Our parametrization has allows for $w(z)$ to have a wide class of behavior in which it  may grow and later decrease or other way around. The EoS  $w(x(y,L))$   in  eq.(\ref{www}) is given in terms of the functions $L$ and $y$  and it is an exact equation, valid also when the slow roll approximation is not satisfied. The EoS $w$ is constrained between $-1\leq w\leq 1$
for any value of $x$, with $0\leq x$ by definition. The  parametrization proposed is given in eqs.(\ref{ll}),
(\ref{yy}) and (\ref{f2}), with $L=L_o+L_1\, y^\xi \,f(z)$  and a transition function $f(z)=\fr{(z/zt)^k}{1+(z/z_t)^k}$,
which has four free parameters: $L_o,L_1, z_t, k$. The parameters $L_o$ and $L_1$ set the values of the EoS at present time $w_o$ and at an early time $w_i=w(z\gg z_t)$, respectively (c.f.  eq.(\ref{lo}) and (\ref{li}). Therefore our EoS has  four free parameters given by   $w_o$ and $w_i$,  the transition redshift  $z_t$ at which the EoS goes from $w_o$ to $w_i$ plus the
the steepness of the transition set by $k$.  Besides studying the evolution of Dark Energy we also determined
its perturbations from  the adiabatic sound speed $c^2_a$ and $c^2_s$ given in eqs.(\ref{ca}) and (\ref{cs}), which are
functions of $x$ and its derivatives.  We have seen that a steep transition has a bump in  $c^2_a$
and this should be detectable in large scale structure if it takes place at late times.

We can use the parametrization of $x(L,y)$ in eqs.(\ref{xl}), (\ref{ll}) and (\ref{yy}) and $c^2_a$ and $c^2_s$
in eqs.(\ref{ca}) and (\ref{cs}) without any reference to the underlying physics, namely the dynamics of the scalar
field $\p$, and the parametrization is well defined.  However, it is when we interpret $x=\dot\p^2/2V$ and $ L= (V'/V)^2 A$ and $y= \rmm/V$ that we are  analyzing the evolution of a scalar field $\p$ and we can  connect the evolution of $w$ to the  potential $V(\p)$, once
the free parameters are phenomenological determined by the cosmological  data.  The slow roll approximation is when
we take $|q|\ll 1, A\simeq 1$.

To conclude,  we have  proposed  a new parametrization of DE  which has a rich structure, and
the determination of its  parameters  will help  us to understand the dynamics of Dark Energy.

\acknowledgments
We acknowledge financial support from  PAPIIT IN101415.

\begin{appendix}

\section{appendix}

The parameter $|q\equiv\ddot \p/V'|$ is clearly smaller than one in the slow roll regime ($\ddot\p < 3H\dot\p\simeq V'$). Let us now determine
the dependence of $q$ on the potential $V(\p)$ and its derivatives.
The evolution of $q$ is
\bea\la{dqq}
\dot q &=& H q_N= \fr{\dddot\p}{V'}-\fr{\ddot\p\dot\p}{V'^2} \\
&=& 3H\le(-q+(1+q)\fr{\dot H}{3H^2}+ 2\G x)\ri)
\eea
where we used eq.(\ref{pq}),
\bea\la{ddd}
\fr{\dddot\p}{V'} &&= - \fr{V''\dot\p}{V'}- \fr{3 \dot H  \dot\p}{V'}-\fr{3H \ddot \p}{V'}\\
&&=-3Hq+3H(1+q)\le(\fr{V''}{9H^2}+\fr{\dot H}{3H^2}\ri),
\eea
and
\be\la{sl3}
\fr{V''}{9H^2}=\fr{\G x }{3},  \;\;\;\;\G\equiv\fr{V''V}{V'^2}.
\ee
We can estimate the value of  $q$
 if  we drop  the term proportional to $\dddot\p$ in  eq.(\ref{ddd})  giving
\be\la{qq}
q \simeq \fr{\fr{V''}{9H^2}+ \fr{\dot H}{3H^2}}{1- (\fr{V''}{9H^2}+ \fr{\dot H}{3H^2})}.
\ee
In a stable evolution of $\p$ we have a positive $V''$ and since $\dot H$ is negative
both terms have opposite signs, but of course we do not expect a complete cancelation of these terms.
However both of them are smaller than one, since
$ 0 \leq - \dot H/3H^2<1/2$  for  $x<1$  and
 $|V''/9H^2|=\G x /3 < 1/3$ in the slow roll approximation.
 A tracker behavior requires $\G>1$  \ci{tracker} and
$x<1/\G<1$. Finally, the evolution of $L$ is given by
$\dot L = H\,L_N= 2 \lm\dot\lm (1+q)^2+\lm^2 \dot q(1+q)=2L[\dot\lm/\lm+\dot q(1+q)]$,
\bea\la{dll}
\dot L  &=&  \fr{12HLx(1-\G)}{(1+q)}+\fr{2L\dot q}{(1+q)}= 3HL \le(\fr{2x-q}{1+q}+\fr{\dot H}{3H^2}  \ri)\nonumber\\
 &=&  3HL\le(\fr{2(q-x)(1+2x+2y)-y(1+q)}{2(1+q)(1+x+y)}\ri).
\eea
We see that at $-1<q\leq x$ we have $\dot L<0$ giving a decreasing $L$ as a function of time or
an increasing $L$ as  a function of  z . For $(q-x)/(1+q)>y/(1+2x+2y) $ or
equivalently for $q > (y+2x(1+2x+2y))/(2+4x+3y)$ we have $\dot L >0$  and
a decreasing $L$ as  a function of  z.

Instead of choosing a DE parametrization as in eq.(\ref{ll}) we could solve eqs.(\ref{dqq}) and (\ref{dll})
for different  potentials $V(\p)$ or by taking different approximated solutions or ansatze for
$\G$. However, we choose to parameterize directly $L$ as in eq.(\ref{ll}). Still
using $\dot L= \dot a L_a= a H L_a$  and $L_1y^\xi f=L-L_o$ we identify
\be
(L-L_o) \le( k\, f \le(\fr{a}{a_t}\ri)^k  - 3\xi  w_o \ri)=  3L \le(\fr{2x-q}{1+q}+\fr{\dot H}{3H^2} \ri)
\ee
and the choices of $\G$ and $q$ would fix the parameters $L_o, k$ and $a_t$.

\end{appendix}

\thebibliography{}

\footnotesize{

 \bib{wmap9}
  G.~Hinshaw, D.~Larson, E.~Komatsu, D.~N.~Spergel, C.~L.~Bennett, J.~Dunkley, M.~R.~Nolta and M.~Halpern {\it et al.},
  arXiv:1212.5226 [astro-ph.CO].

 \bib{planck}
  P.~A.~R.~Ade {\it et al.} [Planck Collaboration],
  arXiv:1502.01589 [astro-ph.CO],
  P.~A.~R.~Ade {\it et al.} [Planck Collaboration],
  arXiv:1502.01590 [astro-ph.CO].
  P.~A.~R.~Ade {\it et al.}  [ Planck Collaboration],
  arXiv:1303.5076 [astro-ph.CO].

\bib{LSS}
  L.~Anderson, E.~Aubourg, S.~Bailey, F.~Beutler, A.~S.~Bolton, J.~Brinkmann, J.~R.~Brownstein and C.~-H.~Chuang {\it et al.},
  arXiv:1303.4666 [astro-ph.CO];

\bib{SN}
  A.~Conley, J.~Guy, M.~Sullivan, N.~Regnault, P.~Astier, C.~Balland, S.~Basa and R.~G.~Carlberg {\it et al.},
  Astrophys.\ J.\ Suppl.\  {\bf 192}, 1 (2011)
  [arXiv:1104.1443 [astro-ph.CO]];
  N.~Suzuki, D.~Rubin, C.~Lidman, G.~Aldering, R.~Amanullah, K.~Barbary, L.~F.~Barrientos and J.~Botyanszki {\it et al.},
  Astrophys.\ J.\  {\bf 746}, 85 (2012)
  [arXiv:1105.3470 [astro-ph.CO]].

\bib{new}
 J. Bock et al. (EPIC Collaboration), arXiv:0906.1188,
BigBOSS Expermient Collaboration [arXiv. 1106.1706];
  ``Euclid Definition Study Report''arXiv:1110.3193 [astro-ph.CO];
  B.~M.~Rossetto {\it et al.}  [Dark Energy Survey Collaboration],
  arXiv:1104.4718 [astro-ph.GA].

\bib{DE.rev}
E. J. Copeland, M. Sami,
and S. Tsujikawa, Int. J. Mod. Phys. D 15, 1753 (2006).

\bib{DEparam}

M. Doran and G. Robbers, J. Cosmol. Astropart. Phys. 06 (2006) 026;
E.V. Linder, Astropart. Phys. 26, 16 (2006);
D. Rubin et al. , Astrophys. J. 695, 391 (2009) [arXiv:0807.1108];
J. Sollerman et al. , Astrophys. J. 703, 1374 (2009)[arXiv:0908.4276];
M.J. Mortonson, W. Hu, D. Huterer, Phys. Rev. D 81, 063007 (2010)  [arXiv:0912.3816];
S. Hannestad, E. Mortsell JCAP 0409 (2004) 001 [astro-ph/0407259];
H.K.Jassal, J.S.Bagla, T.Padmanabhan, Mon.Not.Roy.Astron.Soc. 356, L11-L16 (2005);
S. Lee,  Phys.Rev.D71, 123528 (2005)

\bib{DEParam.Recosntr}
J.~Z.~Ma and X.~Zhang,
  Phys.\ Lett.\  B {\bf 699} (2011) 233 [arXiv:1102.2671];
Dragan Huterer, Michael S. Turner  Phys.Rev.D64:123527 (2001)   [astro-ph/0012510];
Jochen Weller (1), Andreas Albrecht, Phys.Rev.D65:103512,2002   [astro-ph/0106079]

\bib{CPL}
M. Chevallier and D. Polarski, Int. J. Mod. Phys. D10,
213 (2001); E. V. Linder, Phys. Rev. Lett. 90, 091301 (2003).

\bib{corasaniti}
P. S. Corasaniti, B. A. Bassett, C. Ungarelli, and E. J. Copeland, Phys. Rev. Lett. 90, 091303 (2003);

\bib{huang}
Z. Huang, J. R. Bond, L.Kofman Astrophys.J.726:64,2011

\bib{SF.Peebles}
B.~Ratra and P.~J.~E.~Peebles,
  Phys.\ Rev.\  D {\bf 37}, 3406 (1988),C.~Wetterich,
  Astron.\ Astrophys.\  {\bf 301}, 321 (1995)
  [arXiv:hep-th/9408025].

\bib{tracker}
P. Steinhardt, L.Wang, I.Zlatev,  Phys.Rev.Lett. 82 (1999) 896;
Phys.Rev.D 59(1999) 123504

\bib{quint.ax}
  A.~de la Macorra and G.~Piccinelli,
  Phys.\ Rev.\  D {\bf 61}, 123503 (2000)
  [arXiv:hep-ph/9909459];
  A.~de la Macorra and C.~Stephan-Otto,
  Phys.\ Rev.\  D {\bf 65}, 083520 (2002)
  [arXiv:astro-ph/0110460].

\bib{GDE.ax}
  A.~de la Macorra,
  Phys.\ Rev.\  D {\bf 72}, 043508 (2005)
  [arXiv:astro-ph/0409523];
  A.~De la Macorra,
  JHEP {\bf 0301}, 033 (2003)
  [arXiv:hep-ph/0111292];
   A.~de la Macorra and C.~Stephan-Otto,
  Phys.\ Rev.\ Lett.\  {\bf 87}, 271301 (2001)
  [arXiv:astro-ph/0106316];

\bib{GDM.ax}
  A.~de la Macorra,
  Phys.\ Lett.\  B {\bf 585}, 17 (2004)
  [arXiv:astro-ph/0212275],A.~de la Macorra,
  Astropart.\ Phys.\  {\bf 33}, 195 (2010)
  [arXiv:0908.0571 [astro-ph.CO]].

\bib{IDE}
 S.~Das, P.~S.~Corasaniti and J.~Khoury,Phys.  Rev. D { 73},
083509 (2006), arXiv:astro-ph/0510628;
A.~de la Macorra,  
 Phys.Rev.D76, 027301 (2007), arXiv:astro-ph/0701635

\bib{IDE.ax}
  A.~de la Macorra,
  JCAP {\bf 0801}, 030 (2008)
  [arXiv:astro-ph/0703702];
  A.~de la Macorra,
  Astropart.\ Phys.\  {\bf 28}, 196 (2007)
  [arXiv:astro-ph/0702239].

\bib{ma} C. Ma and E. Bertschinger, Astrophys. J. 455, 7 (1995)

\bib{mukhanov}
J. Garriga and V. Mukhanov, Phys. Lett. B 458, 219 (1999)

\bib{hu}
W. Hu, D. Scott, N. Sugiyama and M. J. White, Phys. Rev. D 52,
5498 (1995), e-Print: astro-ph/9505043. W. Hu, Astrophys. J. 506
(1998) 485 [arXiv:astro-ph/9801234]. S. Bashinsky and U. Seljak,
Phys. Rev. D 69, 083002 (2004)

\bib{bean}
  R.~Bean, O.~Dore and ,
  Phys.\ Rev.\ D {\bf 69}, 083503 (2004)
  [astro-ph/0307100].

\bib{DE.p.sf}
  G.~Ballesteros, A.~Riotto and ,
  Phys.\ Lett.\ B {\bf 668}, 171 (2008)
  [arXiv:0807.3343 .
  L.~R.~Abramo, R.~C.~Batista, L.~Liberato, R.~Rosenfeld and ,
  JCAP {\bf 0711}, 012 (2007)
  [arXiv:0707.2882].

\bib{DE.p}
  J.~Liu, M.~Li, X.~Zhang and ,
  JCAP {\bf 1106}, 028 (2011)
  [arXiv:1011.6146 ].
  U.~Alam,
  Astrophys.\ J.\  {\bf 714}, 1460 (2010)
  [arXiv:1003.1259].
  R.~de Putter, D.~Huterer, E.~V.~Linder
  Phys.\ Rev.\ D {\bf 81}, 103513 (2010)
  [arXiv:1002.1311].
  H.~K.~Jassal,
  Phys.\ Rev.\ D {\bf 81}, 083513 (2010)
  [arXiv:0910.1906].
}

\end{document}